# Small-volume effect enables the spine robust, sensitive and efficient information transfer

Masashi Fujii,[1] Kaoru Ohashi,[1] Yasuaki Karasawa,[2] Minori Hikichi,[1] and Shinya Kuroda [1,3,*]

[1]*Department of Biological Sciences, Graduate School of Sciences, University of Tokyo, Hongo 7-3-1, Bunkyo-ku, Tokyo, 113-0033, Japan*

[2]*Department of Neurosurgery, Graduate School of Medicine, University of Tokyo, Hongo 7-3-1, Bunkyo-ku, Tokyo 113-0033, Japan*

[3]*CREST, Japan Science and Technology Agency, Hongo 7-3-1, Bunkyo-ku, Tokyo 113-0033, Japan*

[*]Corresponding author: skuroda@bs.s.u-tokyo.ac.jp

**Abstract**

Why is the spine of a neuron so small that only small numbers of molecules can exist and reactions inevitably become stochastic? Despite such noisy conditions, we previously showed that the spine exhibits robust, sensitive and efficient features of information transfer using probability of $Ca^{2+}$ increase; however, their mechanisms remains unknown. Here we show that the small-volume effect enables robust, sensitive and efficient information transfer in the spine volume, but not in the cell volume. In the spine volume, intrinsic noise in reactions becomes larger than extrinsic noise of input, making robust information transfer against input fluctuation. Stochastic facilitation of $Ca^{2+}$ increase occurs in the spine volume, making higher sensitivity to lower intensity of input. Volume-dependency of information transfer enables efficient information transfer per input in the spine volume. Thus, we propose that the small-volume effect is the functional reasons why the spine has to be so small.

The spine is a platform of a neuron where input timing information from other neurons is integrated, and is extremely small [1,2]. For example, in a parallel fiber (PF)-cerebellar Purkinje cell synapse, the volume of spine is approximately 0.1 μm$^3$ where the number of molecules are limited to tens to hundreds (Fig. 1a, see also Fig. 7), and is 10$^4$-fold smaller than the cell body (5,000 μm$^3$) (Fig. 1a, see also Fig 7) [3–5]. Under such conditions, reactions in the spine inevitably become stochastic and inputs are fluctuated due to low numbers of molecules in the small volume [6–10]. Intuitively, such noisy conditions are disadvantageous for information processing. Why is the spine so small? This is one of the fundamental questions in neuroscience and biological information processing.

Cerebellar Purkinje cells receive two inputs. One is PF inputs from granular neurons that are thought to code sensorimotor signals, and the other is a climbing fiber (CF) input from inferior olivary nucleus that is thought to code error signal [11–13]. Conjunctive PF and CF inputs but not either PF or CF inputs alone has been shown to induce large $Ca^{2+}$ increases by positive feedback loop via $IP_3$ (inositol trisphosphate)-induced $Ca^{2+}$ release (IICR) [14,15], leading to long-term decreases of synaptic strength that are known as cerebellar long-term depression (LTD) [16], a tentative molecular basis of cerebellar motor learning [17,18]. It has experimentally been shown that large $Ca^{2+}$ increase occur when PF and CF inputs are coincident at a given synapse within a 200 msec time window, and that $Ca^{2+}$ increase against the timing between PF and CF inputs shows a bell-shaped response (Fig. 1c) [15]. We have previously developed detailed biochemical deterministic model of $Ca^{2+}$ increase in a PF-cerebellar Purkinje cell synapse, and reproduced the PF- and CF-timing dependent $Ca^{2+}$ increase [19]. In addition, by reducing this model, we have also made the simple deterministic model and extracted essential framework of the network for PF and CF inputs dependent $Ca^{2+}$ increase [20,21].

However, in the spine, the number of molecules is limited to tens to hundreds; thus, reactions should behave stochastically rather than deterministically. Indeed, it has experimentally been shown that, the coincident PF and CF inputs stochastically induce $Ca^{2+}$ increases; in some cases, large $Ca^{2+}$ increases are observed, whereas in other cases, they are not (Fig. 1b, c). In addition to the intrinsic noise due to the stochastic fluctuation of $Ca^{2+}$ increase, PF inputs has been shown to fluctuate [8–10], which can be regarded as extrinsic noise. We have created a stochastic simulation model based on the deterministic model [19] that incorporated stochastic reactions due to the small number of molecules of the interval between PF and CF inputs [22]. We have previously shown that the spine uses probability of $Ca^{2+}$ increase, rather than its amplitude, for information transfer, and that probability of $Ca^{2+}$ increases in the spine shows robustness against input fluctuation, sensitivity to lower input numbers, and efficiency of information transfer [22]. However, the robustness, sensitivity and efficiency were not defined, and their mechanism remains unknown.

In this study, we made a simple stochastic model based on the simple deterministic model [20]. Using the simple stochastic model, we defined the robustness, sensitivity and efficiency of information transfer by $Ca^{2+}$ increase, and clarified their mechanisms (see Fig. 7). Furthermore, we found that the small-volume effect enables robust, sensitive and efficient information transfer in the spine volume, but not in the cell volume. We propose that the small-volume effect is one of the functional reasons why the spine has to be so small.

**Results**

**Development of the simple stochastic model**

To reduce complexity and computational cost of the detailed stochastic model (Fig. 1d) [22], we constructed the simple stochastic model based on the simple deterministic model (Fig. 1e) [20]. We set the parameters regarding to PF and CF inputs of simple deterministic model to reproduces the PF- and CF-timing dependent $Ca^{2+}$ response of the detailed stochastic model [22] (Fig. 1, Supplementary Fig. 1, Supplementary Table 1, 2, Method). Hereafter, we denoted $10^{-1}$ μm$^3$ as the spine volume, and $10^3$ μm$^3$ as the cell volume. In the spine volume, coincident PF and CF inputs with $\Delta t = 100$ msec induced a large $Ca^{2+}$ increase (Fig. 1f, red), but sometimes failed to induce a large $Ca^{2+}$ increase (Fig. 1f, blue). We defined $Ca_{res}$ as the temporal integration of $Ca^{2+}$ concentration, subtracted by the basal $Ca^{2+}$ concentration. The distribution of $Ca_{res}$ showed a bimodal distribution (Fig. 1g). The distribution of $Ca_{res}$ always showed a bimodal distribution regardless of the timing between PF and CF inputs, and probability of a large $Ca^{2+}$ increase changed depending the timing between PF and CF inputs (Fig. 1h). We divided the distribution of $Ca_{res}$ into the probability component (Fig. 1i) and the amplitude component (Fig. 1j) (see Methods). The probability component, but not the amplitude component, showed a bell-shaped time window, indicating that the timing information between PF and CF inputs is coded by the probability of large $Ca^{2+}$ increase, rather than the amplitude of $Ca^{2+}$ increase in the spine volume. By contrast, in the cell volume, the coincident PF and CF inputs with $\Delta t = 100$ msec always induced a large $Ca^{2+}$ increase without failure (Fig. 1k) and $Ca_{res}$ showed a unimodal distribution (Fig. 1l) (see Methods). The distribution of $Ca_{res}$ always showed a unimodal distribution regardless of the timing between PF and CF inputs (Fig. 1m), and only the amplitude of $Ca_{res}$ (Fig. 1o), but not the probability (Fig. 1n), showed a bell-shaped time window, indicating that the timing information between PF and CF inputs is coded by the amplitude of $Ca^{2+}$ increase, rather than the probability of large $Ca^{2+}$ increase in the cell volume. These results are consistent with our previous study using the detailed stochastic model [22].

The simple stochastic model also showed the similar properties, such as efficiency, robustness, and sensitivity in the detailed stochastic model (Fig. 2, Supplementary Fig. 2). The mutual information between $Ca_{res}$ and the PF- and CF-timing increased with the increase in the volume (Fig. 2a). In the spine volume, the probability component of the mutual information was larger than the amplitude component of the mutual information (Fig. 2a, inset), and the amplitude component of the mutual information became larger than the probability component of the mutual information with the increase in the volume. Mutual information per volume became the highest at the spine volume, and decreased with the increase in the volume (Fig. 2b), indicating that the most efficient information coding per volume is achieved at the spine volume. In the spine volume, the mutual information did not decrease and remained constant regardless of the CV (coefficient of variation) of PF input (Fig. 2c, black line), whereas that in the cell volume decreased with the increase in CV of PF input (Fig. 2c, yellow line, Supplementary Fig. 1), indicating that the information transfer by $Ca_{res}$ robust against fluctuation of PF input only in the spine volume, but not in the larger volume including the cell volume. The detailed stochastic model showed a higher sensitivity to the lower numbers of PF input in the spine volume rather than the cell volume (Supplementary Fig. 2d)[22]. We showed that the higher sensitivity to low PF input can be seen in the spine volume, but not in the larger volume including the cell volume (see below). These results in the simple stochastic model are also consistent with those in the detailed stochastic model (Supplementary Fig. 2)[22].

These results indicate that the simple stochastic model can retain the essential properties of Ca$^{2+}$ response, such as robust, sensitive and efficient features. Using this simple stochastic model, we next defined the robustness, sensitivity, and efficiency, and clarified their mechanisms in the spine volume.

**The mechanism of the robustness**

In this section, we defined the robustness, and clarified the mechanism of the robustness as follows. The amplitudes of Ca$^{2+}$ increase by conjunctive PF and CF inputs is compatible with those by strong PF input alone (see Supplementary Fig. 3k)[19]. Consistently, PF input alone has experimentally been shown to induce a large Ca$^{2+}$ increase [23]. Therefore, we hereafter used PF input alone. First, we showed that robustness is given by unchanging of the distribution of $Ca_{res}$ against the fluctuation of PF input. We obtained the necessary and sufficient condition for the robustness, that the intrinsic noise is much larger than the extrinsic noise. We showed that the range of the fluctuation of PF input satisfying the conditions for the robustness is much larger in the spine volume than in the cell volume, indicating that the distribution of $Ca_{res}$ against the fluctuation of PF input in the spine volume is more robust than the cell volume against the fluctuation of PF input.

Hereafter, for simplicity, we used only PF input alone instead of PF and CF inputs. The $Amp_{PF}$, amplitude of the PF input, was set to be between 150 and 215 unless otherwise noted. Consistently, PF input alone has experimentally been shown to induce a large Ca$^{2+}$ increase [23]. We performed the stochastic simulation 10$^4$ times per each amplitude of PF input, which is defined as $Amp_{PF}$, and obtained $p_c(Ca_{res}|Amp_{PF})$, the probability density distribution of $Ca_{res}$. Using $p_c(Ca_{res}|Amp_{PF})$, we examined the mechanism of the robustness. In the spine volume, the distribution of $Ca_{res}$, $p_c(Ca_{res}|Amp_{PF})$, became bimodal when $Amp_{PF}$ exceeded around 50 (Fig. 3a, Supplementary Fig. 3a, b). By contrast, in the cell volume, the distribution of $Ca_{res}$ always showed unimodal distribution regardless of $Amp_{PF}$, and its average monotonically increased along $Amp_{PF}$ when $Amp_{PF}$ exceeded around 150 (Fig. 3b, Supplementary Fig. 3i, j).

*The robustness is given by unchanging of the distribution of $Ca_{res}$ against the fluctuation of PF input*

We have shown that the distribution of $Ca_{res}$ for each $\Delta t$ was unchanged regardless of CV of PF input in the spine volume, but not in the cell volume (see Supplementary Fig. 1) [22], suggesting that unchanging of the distribution of $Ca_{res}$ against the fluctuation of PF input is a key to the robustness of the information transfer. Therefore, we examined whether the distribution of $Ca_{res}$ with PF input alone is also unchanged regardless of CV of PF input in the spine volume, but not in the cell volume.

Experimentally, it has been reported that the distribution of amplitude of PF input in the Purkinje cell can be approximated by a Gaussian distribution [10]. We set $p_c(Amp_{PF}|\mu_a, \sigma_a)$, the probability density distribution of $Amp_{PF}$, as the Gaussian distribution given by $\mathcal{N}(Amp_{PF}|\mu_a, \sigma_a^2)$, where $\mu_a$ and $\sigma_a$ denote the average of $Amp_{PF}$ and the standard deviation of $Amp_{PF}$, respectively. We used $\sigma_a$, the standard deviation of $Amp_{PF}$, as the magnitude of fluctuation of $Amp_{PF}$ because the $\sigma_a$ is proportional to $CV_a$, CV of $Amp_{PF}$, with the fixed $\mu_a$, given by $CV_a = \sigma_a/\mu_a$. When PF input is given by $p_a(Amp_{PF}|\mu_a, \sigma_a)$, $p_{ac}(Ca_{res}|\mu_a, \sigma_a)$, the distribution of $Ca_{res}$ with the fluctuation of $Amp_{PF}$, is given by

$$p_{ac}(Ca_{res}|\mu_a, \sigma_a) = \int_{Amp_{PF}} p_a(a|\mu_a, \sigma_a) p_c(Ca_{res}|a)\, da$$

$$= \int_{Amp_{PF}} \mathcal{N}(a|\mu_a, \sigma_a^2) p_c(Ca_{res}|a)\, da, \qquad (1)$$

where, $p_c(Ca_{res}|a)$ for each $a \in Amp_{PF}$, i.e. $p_c(Ca_{res}|Amp_{PF})$, was obtained by the stochastic simulation. In the spine volume, the distributions of $Ca_{res}$ always exhibited the similar bimodal distributions regardless of $CV_a$ and did not change even if the $CV_a$ became larger (Fig. 3c–e). By contrast, in the cell volume, the distributions of $Ca_{res}$ exhibited unimodal distribution with $CV_a = 0$, and with the increase in $CV_a$, the distributions of $Ca_{res}$ changed and became bimodal (Fig. 3f–h). These properties remained the same regardless of the average of $Amp_{PF}$, $\mu_a$ (Supplementary Fig. 4). The similar results were obtained when both PF and CF inputs were used (see Supplementary Fig. 1).

Taken together, in the spine volume, the distribution of $Ca_{res}$ remained almost unchanged against the fluctuation of $Amp_{PF}$, whereas, in the cell volume, the distribution of $Ca_{res}$ largely varied against the fluctuation of $Amp_{PF}$. These results indicate that unchanging of the distribution of $Ca_{res}$ against the fluctuation of $Amp_{PF}$ causes the robustness.

We quantitated the change of the distributions of $Ca_{res}$ with the increase in $CV_a$ by the $\chi^2$ distance against the distributions of $Ca_{res}$ with $CV_a = 0$. The $\chi^2$ distance becomes 0 when the distribution of $Ca_{res}$ with a $CV_a$ is the same with that with $CV_a = 0$, and the $\chi^2$ distance becomes 1 when two distributions are completely different. In the spine volume, the $\chi^2$ distance remained almost 0 regardless of $CV_a$, whereas, in the cell volume, the $\chi^2$ distance abruptly increased with the increase in $CV_a$ and became close to 1 (Fig. 3i), indicating that the distribution of $Ca_{res}$ in the spine volume unchanges with the increase in $CV_a$, whereas, that in the cell volume largely changes even with the small increase in $CV_a$. This is the reason why the robustness can be seen only in the spine volume but not in the cell volume.

*The necessary and sufficient condition for the robustness*

Next, we clarified the mechanism that the distribution of $Ca_{res}$ in the spine volume unchanges with the increase in $CV_a$, whereas that in the cell volume largely changes even with the small increase in $CV_a$, and obtained the necessary and sufficient conditions for the robustness: unchanging of the distribution of the distribution of $Ca_{res}$ regardless of $CV_a$. We considered $p_{ac}(Ca_{res}|\mu_a, \sigma_a)$, the distribution of $Ca_{res}$ with the fluctuation of $Amp_{PF}$, with the increase in $CV_a$. Note that $CV_a = \sigma_a/\mu_a$. Since $\mathcal{N}(Amp_{PF}|\mu_a, \sigma_a^2)$ is symmetric with respect to the point $\mu_a$, *i.e.* $\forall a \in Amp_{PF}$, $\mathcal{N}(a|\mu_a, \sigma_a^2) = \mathcal{N}(2\mu_a - a|\mu_a, \sigma_a^2)$, the equation (1) was deformed into

$$p_{ac}(Ca_{res}|\mu_a, \sigma_a) = \int_{\mu_a}^{\infty} \mathcal{N}(a|\mu_a, \sigma_a^2) p_c(Ca_{res}|a)\, da + \int_{-\infty}^{\mu_a} \mathcal{N}(a|\mu_a, \sigma_a^2) p_c(Ca_{res}|a)\, da$$

$$= \int_{\mu_a}^{\infty} \mathcal{N}(a|\mu_a, \sigma_a^2) p_c(Ca_{res}|a)\, da + \int_{\mu_a}^{\infty} \mathcal{N}(2\mu_a - a|\mu_a, \sigma_a^2) p_c(Ca_{res}|2\mu_a - a)\, da$$

$$= \int_{\mu_a}^{\infty} [\mathcal{N}(a|\mu_a, \sigma_a^2) p_c(Ca_{res}|a) + \mathcal{N}(a|\mu_a, \sigma_a^2) p_c(Ca_{res}|2\mu_a - a)]\, da$$

$$= \int_{\mu_a}^{\infty} \mathcal{N}(a|\mu_a, \sigma_a^2) [p_c(Ca_{res}|a) + p_c(Ca_{res}|2\mu_a - a)]\, da. \tag{2}$$

Since the distribution of $Amp_{PF}$ is the Gaussian distribution, $\mathcal{N}(Amp_{PF}|\mu_a, \sigma_a^2)$, the probability density that the $Amp_{PF} = a$ occurs decreases as the difference between $a$ and $\mu_a$ becomes larger. In particular, the probability that $Amp_{PF} = a$ is included in the range $\mu_a - 3\sigma_a \leq a \leq \mu_a + 3\sigma_a$ is given by $\int_{\mu_a - 3\sigma_a}^{\mu_a + 3\sigma_a} \mathcal{N}(a|\mu_a, \sigma_a^2)\, da = 0.9974...$, *i.e.* almost 1. Thus, the probability of $a > \mu_a + 3\sigma_a$ or $a < \mu_a - 3\sigma_a$ is quite small and negligible. Therefore, satisfying the equation (2) in the range $\mu_a - 3\sigma_a \leq a \leq \mu_a + 3\sigma_a$ is enough to satisfy the equation (2) in almost all the range of $Amp_{PF}$. This means the r.h.s. (right-hand side) of the equation (2) in the range $a > \mu_a + 3\sigma_a$ can be neglected, and $p_{ac}(Ca_{res}|\mu_a, \sigma_a)$ is given by

$$p_{ac}(Ca_{res}|\mu_a, \sigma_a) = \int_{\mu_a}^{\mu_a + 3\sigma_a} \mathcal{N}(a|\mu_a, \sigma_a^2) [p_c(Ca_{res}|a) + p_c(Ca_{res}|2\mu_a - a)]\, da. \tag{3}$$

Here, we considered the two alternative cases. One is the case where the averaged distribution between the distributions of $Ca_{res}$ with the $Amp_{PF} = a$ shifted $|a - \mu_a|$ from $\mu_a$, the average of $Amp_{PF}$, *i.e.*, $1/2[p_c(Ca_{res}|a) + p_c(Ca_{res}|2\mu_a - a)]$, is almost the same as the distribution of $Ca_{res}$ with $Amp_{PF} = \mu_a$, $p_c(Ca_{res}|\mu_a)$ up to $a = \mu_a + 3\sigma_a$, given by

$$\forall a < \mu_a + 3\sigma_a:$$

$$p_c(Ca_{res}|\mu_a) \simeq \frac{1}{2}[p_c(Ca_{res}|a) + p_c(Ca_{res}|2\mu_a - a)]. \tag{4}$$

The other is the case where the averaged distribution of $Ca_{res}$ between the distributions of $Ca_{res}$ with $Amp_{PF}$ shifted $|a - \mu_a|$ from $\mu_a$ is not always the same as the distribution of $Ca_{res}$ with $\mu_a$ up to $a = \mu_a + 3\sigma_a$, given by

$$\exists a < \mu_a + 3\sigma_a:$$

$$p_c(Ca_{res}|\mu_a) \not\simeq \frac{1}{2}[p_c(Ca_{res}|a) + p_c(Ca_{res}|2\mu_a - a)]. \tag{5}$$

First, we considered the conditions where the equation (4) is satisfied, and obtained the condition where $p_{ac}(Ca_{res}|\mu_a, \sigma_a)$, the distribution of $Ca_{res}$ with the fluctuation of $Amp_{PF}$, does not change against $\sigma_a$, the magnitude of fluctuation of $Amp_{PF}$. This condition means that the distribution of $Ca_{res}$ remained the same against fluctuation of $Amp_{PF}$. Substituting the equation (4) for the equation (3), we obtained

$$p_{ac}(Ca_{res}|\mu_a, \sigma_a) \simeq 2 \int_{\mu_a}^{\mu_a + 3\sigma_a} \mathcal{N}(a|\mu_a, \sigma_a^2) p_c(Ca_{res}|\mu_a)\, da$$

$$= 2 p_c(Ca_{res}|\mu_a) \int_{\mu_a}^{\mu_a + 3\sigma_a} \mathcal{N}(a|\mu_a, \sigma_a^2)\, da$$

$$\simeq p_c(Ca_{res}|\mu_a). \tag{6}$$

The l.h.s. (left-hand side) of the equation (6) indicates the distribution of $Ca_{res}$ with the fluctuation of $Amp_{PF}$. The l.h.s. is almost the same as the r.h.s. that indicates the distribution of $Ca_{res}$ without the fluctuation of $Amp_{PF}$. Note that the $\sigma_a$ does not directly appear in the equation (4), however, $\sigma_a$ determines the upper bound of the range of $a - \mu_a$ satisfying the equation (4). This means that if the equation (6) is satisfied for $\sigma_a = \sigma_a^*$, the equation (6) is also satisfied for $\sigma_a < \sigma_a^*$. Namely, the upper bound of the range of $a - \mu_a$ satisfying the equation (4) is larger, the equation (6) is satisfied for larger $\sigma_a$, i.e. $CV_a$. Therefore, the equation (4) is the condition sufficient to allow that the distribution of $Ca_{res}$ does not change against the fluctuation of $Amp_{PF}$.

By contrast, under the conditions where the equation (5) is satisfied, there does not exist $p_{ac}(Ca_{res}|\mu_a, \sigma_a)$, the distribution of $Ca_{res}$ with the fluctuation of $Amp_{PF}$, which does not change against the fluctuation of $Amp_{PF}$ (see Supplementary Note 1).

Taken together, the equation (4) is a necessary and sufficient condition where the distribution of $Ca_{res}$ remains the same against the fluctuation of $Amp_{PF}$. If $a - \mu_a$, the effective $Amp_{PF}$ with the fluctuation of $Amp_{PF}$, satisfying the equation (4) is larger, the distribution of $Ca_{res}$ does not change even against the larger fluctuation of $Amp_{PF}$. Therefore, the upper bound of the range of $a - \mu_a$ satisfying the equation (4) determines the maximum of $\sigma_a$ where the distribution of $Ca_{res}$ does not change. Next, we examined the upper bounds of the range of $a - \mu_a$ satisfying the equation (4) in the spine volume and in the cell volume. We also demonstrated that the upper bound of the range of $a - \mu_a$ satisfying the equation (4) in the spine volume is much larger than that in the cell volume, thus, information transfer by $Ca_{res}$ in the spine volume is much more robust than that in the cell volume against the fluctuation of $Amp_{PF}$.

*The necessary and sufficient condition for the robustness is satisfied in the range where the intrinsic noise is larger than the extrinsic noise*

We next showed that the upper bound of the range of $Amp_{PF}$ satisfying the equation (4) is determined by the upper bound of the range of $Amp_{PF}$ where the intrinsic noise is larger than the extrinsic noise. We first gave the intuitive interpretation of this proposition using schematic representation of the distribution of $Ca_{res}$ with indicated $Amp_{PF}$ in the spine volume and cell volume (Fig. 3j, Supplementary Fig. 3) and then proved it. The distribution of $Ca_{res}$ in the spine volume is divided into two distributions by threshold $\theta$ (Fig. 3j, see Supplementary Fig. 3). Note that because of the unimodal distribution of $Ca_{res}$ in the cell volume, we set $\theta = -\infty$ in the cell volume. This means that the equation (4) was divided into the forms given by

$$p_c(Ca_{res}|Amp_{PF}) = P_+(Amp_{PF})p_c(Ca_{res}|Ca_{res} > \theta, Amp_{PF})$$
$$+ P_-(Amp_{PF})p_c(Ca_{ress}|Ca_{res} \leq \theta, Amp_{PF}), \tag{7}$$

$$\begin{cases} P_+(Amp_{PF}) \equiv \int_\theta^\infty p_c(c|Amp_{PF})\,dc \\ P_-(Amp_{PF}) \equiv \int_{-\infty}^\theta p_c(c|Amp_{PF})\,dc \end{cases}, \tag{8}$$

where $P_+$ and $P_-$ denote the probabilities of $Ca_{res} > \theta$ and $Ca_{res} \leq \theta$ with $Amp_{PF} = a$, respectively. We separately considered the first term and second term of r.h.s. It should be noted that since $\theta$ in the cell volume was set at $-\infty$, $P_+$ in the cell volume is always 1 for any $Amp_{PF}$. Furthermore, we defined $Amp'_{PF} \equiv Amp_{PF} - \mu_a$, the relative amplitude of PF input, as the difference of $Amp_{PF}$ from $\mu_a$, the average of $Amp_{PF}$. $\hat{p}_c(Ca_{res}|x)$, the distribution of $Ca_{res}$ with $Amp'_{PF} = x$, was defined from $p_c(Ca_{res}|a)$, the distribution of $Ca_{res}$ with $Amp_{PF} = a$, given by

$$\hat{p}_c(Ca_{res}|x) \equiv p_c(Ca_{res}|\mu_a + x) = p_c(Ca_{res}|a) \tag{9}$$

Then, the equations (4) and (7) were deformed, respectively, given by

$$\hat{p}_c(Ca_{res}|0) = \frac{1}{2}[\hat{p}_c(Ca_{res}|x) + \hat{p}_c(Ca_{res}|-x)], \tag{10}$$

$$\hat{p}_c(Ca_{res}|x) = P_+(\mu_a + x)\hat{p}_c(Ca_{ress}|Ca_{res} > \theta, x) + P_-(\mu_a + x)\hat{p}_c(Ca_{ress}|Ca_{res} \leq \theta, x). \tag{11}$$

In the spine volume, the distributions of $Ca_{res}$ above the threshold $\theta$ with $Amp'_{PF} = x$ (Fig. 3j, blue in the left panel, blue line) and with $Amp'_{PF} = -x$ (Fig. 3j, the left panel, the red line) had $\sigma_c(\mu_a \pm x)$, the standard deviation of $Ca_{res}$, which is larger than $\Delta Ca^*$, the gap of the $Ca_{res}$ realizing the peaks of distributions of $Ca_{res}$, and these distributions widely overlapped each other. Then, the averaged distribution of these two distributions of $Ca_{res}$ with $Amp'_{PF} = \pm x$ became the unimodal and intermediate distribution (Fig. 3j, the left panel, the green dashed line), and became almost the same as the distribution of $Ca_{res}$ above threshold $\theta$ with $Amp'_{PF} = 0$ (Fig. 3j, black line in the left panel, the black line). Also, the distributions of $Ca_{res}$ below the threshold $\theta$ exhibited the similar unimodal distribution. Thus, the averaged distribution of these two distributions below the threshold $\theta$ (Fig. 3j, the left panel, the green dashed line) became almost the same as the distribution of $Ca_{res}$ below the threshold $\theta$ with $Amp'_{PF} = 0$ (Fig. 3j, the left panel, the black line). Therefore, in the spine volume, for the both distributions of $Ca_{res}$ above and below the threshold $\theta$, the averaged distributions of the distributions of $Ca_{res}$ with $Amp'_{PF} = \pm x$ were the same as that with $Amp'_{PF} = 0$, indicating that the equation (4) is satisfied. This also means that any symmetrical distribution of $Amp'_{PF}$ other than the Gaussian distribution can give the same result. By contrast, in the cell volume, the distributions of $Ca_{res}$ with $Amp'_{PF} = x$ (Fig. 3j, the right panel, the red line) and $Amp'_{PF} = -x$ (Fig. 3j, the right panel, the blue line) had the standard deviations which are smaller than $\Delta Ca^*$, the gap of the $Ca_{res}$ realizing the peaks of distributions of $Ca_{res}$, and did not overlapped each other. Then, the averaged distribution of these two distributions (Fig. 3j, the right panel, the green dashed line) became bimodal and did not conform to the distribution of $Ca_{res}$ with $Amp'_{PF} = 0$ (Fig. 3j, the right panel, the black line), indicating that the equation (4) is not satisfied. Therefore, the symmetry of the distribution of $Amp'_{PF}$ and large $\sigma_c$, the standard deviations of the distribution of $Ca_{res}$, in comparison to $\Delta Ca^*$, the gap of the $Ca_{res}$ realizing peaks of distributions of $Ca_{res}$, can give the conformation of the averaged distribution of the two distributions of $Ca_{res}$ with $Amp'_{PF} = \pm x$ to the distribution with $Amp'_{PF} = 0$. Then, we proved this proposition. For this purpose, we derived the upper bound of the range of $x$ where the equation (4) is satisfied, and showed that this upper bound in the spine volume is larger than that in the cell volume.

We tried to examine the upper bound of the range of $x$ where the equation (4) is satisfied, and showed that the upper bound of the range of $x$ in the spine volume is larger than that in the cell volume. Hereafter, each distribution of $Ca_{res}$ for $Ca_{res} > \theta$ and $Ca_{res} \leq \theta$ is approximated by the Gaussian distribution. We tried to examine that the equation (4) is satisfied when $\sigma_c$, the standard deviation of $Ca_{res}$, is larger than $\Delta Ca^*$, the gap of $Ca^*$ with $Amp'_{PF} = x$ and $Amp'_{PF} = -x$.

Here, we considered the small gap of $Amp'_{PF}$, hence, for simplicity, $\sigma_c(\mu_a + x)$ and $\sigma_c(\mu_a - x)$, the standard deviations of $Ca_{res}$ with $Amp_{PF} = \mu_a + x$ and $Amp'_{PF} = \mu_a - x$, were regarded as $\sigma_c(\mu_a)$, the standard deviation of $Ca_{res}$ with $Amp_{PF} = \mu_a$, up to the upper bound of the range of $x$ satisfying the equation (4) (Supplementary Fig. 5).

First, we considered $\hat{p}_c(Ca_{res}|Ca_{res} > \theta, x)$, the distribution of $Ca_{res}$, for $Ca_{res} > \theta$ in the spine and cell volumes, we approximated the distribution of $Ca_{res}$ for $Ca_{res} > \theta$ by the Gaussian distribution, given by

$$\hat{p}_c(Ca_{res}|Ca_{res} > \theta, x) \simeq \frac{1}{\sqrt{2\pi\sigma_c^2}} \exp\left[-\frac{(Ca_{res} - Ca^*(\mu_a + x))^2}{2\sigma_c^2}\right]. \tag{12}$$

$Ca^*$ indicates the $Ca_{res}$ realizing the peak of distribution of $Ca_{res}$, given by

$$Ca^*(a) = \arg\max_{Ca_{res}} p_c(Ca_{res}|Ca_{res} > \theta, a). \tag{13}$$

As mentioned above, we assumed $\sigma_c \equiv \sigma_c(\mu_a \pm x) = \sigma_c(\mu_a)$.

Then, for $Ca_{res} > \theta$, we substituted the equations (11) and (12) into the r.h.s. of the equation (10), and obtained

$$\frac{1}{2}[\hat{p}_c(Ca_{res}|x) + \hat{p}_c(Ca_{res}|-x)]$$

$$\simeq \frac{1}{2}\left\{\frac{P_+(\mu_a + x)}{\sqrt{2\pi\sigma_c^2}} \exp\left[-\frac{(Ca_{res} - Ca^*(\mu_a + x))^2}{2\sigma_c^2}\right] + \frac{P_+(\mu_a - x)}{\sqrt{2\pi\sigma_c^2}} \exp\left[-\frac{(Ca_{res} - Ca^*(\mu_a - x))^2}{2\sigma_c^2}\right]\right\}. \tag{14}$$

Here, we considered $Ca^*$. $Ca^*$ for $Ca_{res} > \theta$ linearly increased from around $Amp_{PF} = 50$ in the spine volume (Fig. 4a, black line). In the spine volume, $Ca^*$ for $Ca_{res} > \theta$ linearly increased with the increase in $Amp_{PF}$ for $150 \leq Amp_{PF} \leq 215$, which corresponds to the range of the PF and CF input timing. Thus, regarding $Ca^*$ for $Ca_{res} > \theta$, we could assume

$$Ca^*(\mu_a \pm x) \simeq Ca^*(\mu_a) \pm \Delta Ca^*(x). \tag{15}$$

Equation (15) indicates that the difference of $Ca^*$ between with $Amp_{PF} = \mu_a + x$ and with $Amp_{PF} = \mu_a$ is the same as that between with $Amp_{PF} = \mu_a$ and with $Amp_{PF} = \mu_a - x$, where $\Delta Ca^*$ indicates the difference of $Ca^*$ between with $Amp_{PF} = \mu_a \pm x$ and with $Amp_{PF} = \mu_a$. By contrast to the spine volume, in the cell volume, $Ca^*$ abruptly arose at $Amp_{PF} = 150$, and gradually increased with the increase in $Amp_{PF}$ (Fig. 4a, yellow line). Therefore, in the cell volume, the equation (15) is not satisfied around $Amp_{PF} = 150$, but almost satisfied for $150 < Amp_{PF} \leq 215$. Then, we substituted the equation (15) into the equation (14), and obtained

$$\simeq \frac{1}{2}\left\{ \frac{P_+(\mu_a + x)}{\sqrt{2\pi\sigma_c^2}} \exp\left[-\frac{(Ca_{res} - Ca^*(\mu_a) - \Delta Ca^*(x))^2}{2\sigma_c^2}\right]\right.$$

$$\left. + \frac{P_+(\mu_a - x)}{\sqrt{2\pi\sigma_c^2}} \exp\left[-\frac{(Ca_{res} - Ca^*(\mu_a) + \Delta Ca^*(x))^2}{2\sigma_c^2}\right]\right\}$$

$$= \frac{1}{2\sqrt{2\pi\sigma_c^2}} \exp\left[-\frac{(Ca_{res} - Ca^*(\mu_a))^2}{2\sigma_c^2}\right] \exp\left[-\frac{\Delta Ca^*(x)^2}{2\sigma_c^2}\right]$$

$$\times \left\{P_+(\mu_a + x)\exp\left[\frac{(Ca_{res} - Ca^*(\mu_a))\Delta Ca^*(x)}{\sigma_c^2}\right] + P_+(\mu_a - x)\exp\left[-\frac{(Ca_{res} - Ca^*(\mu_a))\Delta Ca^*(x)}{\sigma_c^2}\right]\right\}. \quad (16)$$

Here, we considered the range of $Ca_{res}$ where $|Ca_{res} - Ca^*(x)| \leq 3\sigma_c(x)$ is almost satisfied. Hence, if $\Delta Ca^*(x) \ll \sigma_c(x)$, then, we could approximate

$$\simeq \frac{1}{\sqrt{2\pi\sigma_c^2}} \exp\left[-\frac{(Ca_{res} - Ca^*(\mu_a))^2}{2\sigma_c^2}\right] \left\{\frac{P_+(\mu_a + x) + P_+(\mu_a - x)}{2}\right\}. \quad (17)$$

Note that, as mentioned below, the upper bound of the range of $x$ where $\Delta Ca^* \ll \sigma_c$ determines the upper bound of the range where the equation (4) is satisfied. This means that the larger upper bound of the range of $x$ where $\Delta Ca^* \ll \sigma_c$ corresponds to the maximum of $CV_a$ with which the distribution of $Ca_{res}$ does not change.

Here, we considered the probability that $Ca_{res}$ exceeds the threshold $\theta$, $P_+$. In the spine volume, $P_+$ gradually arose from $Amp_{PF} = 50$, linearly increased for $100 \leq Amp_{PF} \leq 250$ (Fig. 4b, black line). Therefore, in the spine volume, $P_+$ linearly increased with the increase in $Amp_{PF}$ for $150 \leq Amp_{PF} \leq 215$, which corresponds to the range of the PF-CF input timing. Thus, regarding $P_+$, we could assume

$$\frac{1}{2}[P_+(\mu_a + x) + P_+(\mu_a - x)] = P_+(\mu_a). \quad (18)$$

This equation indicates that the average of the probabilities that $Ca_{res}$ exceeds the threshold $\theta$ with $Amp_{PF} = \mu_a + x$ and $Amp_{PF} = \mu_a - x$ is the same as the probability that $Ca_{res}$ exceeds the threshold $\theta$ with $Amp_{PF} = \mu_a$. In the cell volume, the distribution of $Ca_{res}$ was unimodal, and $\theta = -\infty$ was assumed, then, $P_+$ was always 1, and the equation (18) was always satisfied. Therefore, we substituted the equation (18) into the equation (17), and obtained

$$\simeq \frac{P_+(\mu_a)}{\sqrt{2\pi\sigma_c^2}} \exp\left[-\frac{(Ca_{res} - Ca^*(\mu_a))^2}{2\sigma_c^2}\right] = \hat{p}_c(Ca_{res}|0). \quad (19)$$

for $Ca_{res} > \theta$, i.e. the equation (4) for $Ca_{res} > \theta$ is satisfied.

On the other hand, for $Ca_{res} \leq \theta$, because $Ca^*$ for $Ca_{res} \leq \theta$ was almost constant, the distribution $Ca_{res}$ was mainly characterized only by $P_-$ of the distribution of $Ca_{res}$, indicating

$$\hat{p}_c(Ca_{res}|Ca_{res} \leq \theta, 0) = \hat{p}_c(Ca_{res}|Ca_{res} \leq \theta, \pm x). \qquad (20)$$

Then, using the equation (4) for $Ca_{res} \leq \theta$, similar to the case for $Ca_{res} > \theta$, we obtained

$$\hat{p}_c(Ca_{res}|, 0) = P(0)\hat{p}_c(Ca_{res}|Ca_{res} \leq \theta, 0)$$

$$\simeq \frac{1}{2}[P_-(x) + P_-(-x)]\hat{p}_c(Ca_{res}|Ca_{res} \leq \theta, 0)$$

$$= \frac{1}{2}[P_-(x)\hat{p}_c(Ca_{res}|Ca_{res} \leq \theta, 0) + P_-(-x)\hat{p}_c(Ca_{res}|Ca_{res} \leq \theta, 0)]$$

$$= \frac{1}{2}[P_-(x)\hat{p}_c(Ca_{res}|Ca_{res} \leq \theta, x) + P_-(-x)\hat{p}_c(Ca_{res}|Ca_{res} \leq \theta, -x)]$$

$$= \frac{1}{2}[\hat{p}_c(Ca_{res}|x) + \hat{p}_c(Ca_{res}|, -x)] \qquad (21)$$

for $Ca_{res} \leq \theta$, i.e. the equation (4) for $Ca_{res} \leq \theta$ is also satisfied. Therefore, from the equations (19) and (21), we derived the equation (4). Thus, we approximately showed that if $Ca^*$ and $P_+$, linearly increase with the increase in $Amp_{PF}$ and $\Delta Ca^* \ll \sigma_c$, the equation (4) was satisfied. This means that the necessary and sufficient condition for the robustness is satisfied in the range where the intrinsic noise, $\sigma_c$, is larger than the extrinsic noise, $\Delta Ca^*$.

*The range of the fluctuation of PF input satisfying the conditions for the robustness is larger in the spine volume than in the cell volume*

We next examined the range of the fluctuation of PF input satisfying the condition for the robustness. In the spine volume, in the range considered ($150 \leq Amp_{PF} \leq 215$), $Ca^*(x)$ and $P_+$ always linearly increase with the increase in $Amp_{PF}$. Thus, the range of $Amp_{PF}$ where the distribution of $Ca_{res}$ remains the same regardless of $CV_a$ in the spine volume was determined by the range satisfying $\Delta Ca^* \ll \sigma_c$. In the spine volume, $\Delta Ca^*/\sigma_c \ll 1$ when $x$ was small, $\Delta Ca^*/\sigma_c$ increased with the increase in $x$, and exceeded 1 at $x = 110$ (Fig. 4c). By contrast, in the cell volume, $\Delta Ca^*/\sigma_c$ exceeded 1 even at $x = 2$ (Fig. 4d). We defined $\delta_{max}$ as $x$ that gives $\Delta Ca^*/\sigma_c = 1$. $\delta_{max}$ relatively provides the upper bound of $x$ where $\Delta Ca^*/\sigma_c \ll 1$. Thus, we used $\delta_{max}$ as the index of the robustness (Fig. 4e). The larger $\delta_{max}$ means the more robustness. $\delta_{max}$ was the highest at the spine volume and decreased with the increase in volume (Fig. 4e), indicating that the spine volume gives the highest robustness. Since $\delta_{max}$ in the spine volume was much larger than that in the cell volume, the upper bound of $x$ where $\Delta Ca^*/\sigma_c \ll 1$ in the spine volume is much larger than that in the cell volume. Here, $x$ denoted the relative amplitude of PF input as the displacement of $Amp_{PF}$ from average of $Amp_{PF}$, $\mu_a$, given by $Amp'_{PF} = Amp_{PF} - \mu_a$, i.e. larger $x$ corresponds to larger $CV_a$. Therefore, in the spine volume, in the range $150 < Amp_{PF} \leq 215$, since $\Delta Ca^*/\sigma_c$ was smaller than 1 even with larger $x$, information transfer by $Ca_{res}$ is robust with larger $CV_a$. By contrast, in the cell volume, since $\Delta Ca^*/\sigma_c$ was larger than 1 even with small $x$, information transfer by $Ca_{res}$ is not robust even with small $CV_a$.

Next, we confirmed that, when $\Delta Ca^*/\sigma_c$ is small than 1, the distribution of $Ca_{res}$ with $Amp_{PF} = \mu_a$ and the averaged distribution of $Ca_{res}$ for $Amp_{PF} = \mu_a \pm x$ becomes the same. We quantified the similarities between the two distributions of $Ca_{res}$ by the $\chi^2$ distance. In the spine volume (Fig. 4f, red), most of $\Delta Ca^*/\sigma_c$ were smaller than 1, and the $\chi^2$ distance were also small, indicating that $\Delta Ca^*/\sigma_c$ was smaller than 1 and the two distributions of $Ca_{res}$ were the quite similar in the spine volume. By contrast, in the cell volume (Fig. 4f, blue), most of $\Delta Ca^*/\sigma_c$ were larger than 1, and the $\chi^2$ distance were almost 1 indicating that $\Delta Ca^*/\sigma_c$ was larger than 1, and the two distributions of $Ca_{res}$ were quite different. Therefore, when $\Delta Ca^*$, the gap of two distributions of $Ca_{res}$, is smaller than $\sigma_c$, the standard deviation of the distribution of $Ca_{res}$, the distribution of $Ca_{res}$ unchanges and becomes robust against fluctuation of PF input.

In summary, in the spine volume, $\sigma_c$, the standard deviation of the distribution of $Ca_{res}$, is larger than $\Delta Ca^*$, the gap of the $Ca_{res}$ realizing the peaks of distributions of $Ca_{res}$, that reflects the fluctuation of $Amp_{PF}$, indicating that the range of $x$ satisfying $\Delta Ca^* \ll \sigma_c$ is wider. This means that the distribution of $Ca_{res}$ with fluctuation of amplitude of PF input conforms to that without fluctuation of amplitude of PF input. Moreover, $\Delta Ca^* \ll \sigma_c$ indicates that the gap of distribution $Ca_{res}$ caused by extrinsic noise, $\Delta Ca^*$, is much smaller than that caused by intrinsic noise, $\sigma_c$. Hence, the information transfer by $Ca_{res}$ becomes robust against fluctuation of amplitude of PF input. By contrast, in the cell volume, the standard deviation of the distribution of $Ca_{res}$ without fluctuation of amplitude of PF input is small, and the averaged distribution of $Ca_{res}$ with fluctuation of amplitude of PF input does not conform to that without fluctuation of amplitude of PF input. Moreover, $\Delta Ca^* > \sigma_c$ indicates that the gap of distribution of $Ca_{res}$ caused by extrinsic noise, $\Delta Ca^*$, is larger than that caused by intrinsic noise, $\sigma_c$. Hence, the information transfer by $Ca_{res}$ is not robust against fluctuation of amplitude of PF input.

**The mechanism of the sensitivity**

In the detailed stochastic model, the $Ca^{2+}$ response was sensitive to lower numbers of PF inputs in the spine volume than the cell volume [22]. We tried to examine the sensitivity in the simple stochastic model, and defined the "sensitivity" as follows. For each volume, the PF input was given by the Gaussian distribution with the fixed standard deviation, and the average amplitude of PF input was varied (see Methods). The average amplitude, $\mu_s$, giving the maximum of mutual information, $Amp^*$, was defined as an index of sensitivity for each volume. Smaller $Amp^*$ indicates higher sensitivity to lower amplitude of PF input.

In the spine volume, the mutual information exhibited the bell-shaped response where $Amp^* = 100$ gives the maximum mutual information (Fig. 5a, black line and the white triangle, Fig. 5b, Supplementary Fig. 6a, f). With the increase in volume, $Amp^*$ shifted to around 220 (Fig. 5a, orange line and the black triangle, Fig. 5b, Supplementary Fig. 6e, j). This result indicates that the spine volume shows the higher sensitivity to the lower amplitude of PF input than larger volume including the cell volume.

We considered the mechanism that the spine shows the higher sensitivity to the lower amplitude of PF input. When the standard deviation of PF input distribution is the same, the mutual information depends on $\Delta Ca_{STD}^*$, the dynamic range of the output, and $\sigma_c$, the standard deviation of the output (Supplementary Fig. 7). For example, when the dynamic range of the output is the same, the smaller standard deviation of the output gives the higher mutual information. When the standard deviation of the output is the same, the broader dynamic range gives the higher mutual information. For simplicity, the window width of the input distribution of $Amp_{PF}$ was set as the finite range defined as the average ± standard deviation of the input distribution of $Amp_{PF}$, i.e. $\mu_s \pm STD$, and the dynamic range was denoted as the gap of $Ca_{res}$ realizing peaks of distributions of $Ca_{res}$ between upper bound ($Amp_{PF} = \mu_s + STD$) and lower bound ($Amp_{PF} = \mu_s - STD$) of the input distribution of $Amp_{PF}$, i.e. $\Delta Ca_{STD}^* = Ca^*(\mu_s + STD) - Ca^*(\mu_s - STD)$.

First we considered $\Delta Ca_{STD}^*$, the dynamic range of the output. We defined $\psi(V)$ for each volume as $Amp_{PF}$ where the $Ca^*$ begins to increase (Supplementary Fig. 8a). In the spine volume, $Ca^*$ linearly increased along $Amp_{PF}$ for $Amp_{PF} > \psi(V)$ (Fig. 4a, Supplementary Fig. 8a), hence, $\Delta Ca_{STD}^*$ was largely variable and independent of $Amp_{PF}$ (Fig. 5c, Supplementary Fig. 8b, e–h). By contrast, in the cell volume, $\Delta Ca_{STD}^*$ was bell-shaped curve with the maximum at $Amp_{PF} = 200$ (Fig. 5d, Supplementary Fig. 8c, u–x).

Next, we considered $\sigma_c$, the standard deviation of $Ca_{res}$ for $Ca_{res} > \theta$. In the spine volume, $\sigma_c$ gradually increased with the increase in $Amp_{PF}$ for $Amp_{PF} > \psi(10^{-1}) \simeq 50$ (Fig. 5c, blue line). By contrast, in the cell volume, $\sigma_c$ became largest around $Amp_{PF} = 150$ and gradually decrease with the increase in $Amp_{PF}$ (Fig. 5d, blue line).

In the spine volume, $\Delta Ca_{STD}^*$ was almost constant for $Amp_{PF} > 60$ and $\sigma_c$ increased along $Amp_{PF}$, and therefore, the mutual information became maximum around $Amp_{PF} = 60$ (see Supplementary Fig. 8e–h, black dashed line, also see Supplementary Fig. 6a, f). By contrast, in the cell volume, the mutual information became maximum at $Amp_{PF} = 235$, which is greater than $Amp_{PF} = 200$ giving the maximum of $\Delta Ca_{STD}^*$ (Fig. 5a, d, see Supplementary Fig. 8u–x, black dashed line, also see Supplementary Fig. 6e, j). This is because despite the higher $\Delta Ca_{STD}^*$, $\sigma_c$ was larger and the loss of information became large. Decreasing $\sigma_c$ resulted in increase of mutual information.

Thus, the mutual information becomes the maximum at $Amp_{PF} = 60$ in the spine volume and at $Amp_{PF} = 250$ in the cell volume, indicating the higher sensitivity to lower amplitude of PF input in the spine volume.

**The mechanism of the efficiency**

We defined the efficiency as the mutual information per PF input. The average of PF input was given by $\mu_s \times V$, whose dimension is equal to number of molecule. The efficiency means how much information can be transferred by a unitary PF input. The higher mutual information per PF input indicates the higher efficiency. The mutual information monotonically increased with the increase in volume, and the rate of the increase of the mutual information decreased with the increase in volume (Fig. 6a, black) and therefore, the mutual information per PF input monotonically decreased (Fig. 6b, black line), indicating that the mutual information per PF input, *i.e.* the efficiency, was larger in the spine volume, and decreased as the volume increased to the cell volume (Fig. 6b, black line).

Next, we examined the mechanism of the volume-dependency of the mutual information. The slope of the mutual information decreased with the increase in volume, and became close to logarithmic increase in the larger volume (Fig. 6a, black). Then, we assumed that the volume-dependent increase of the mutual information is roughly approximated with constants, $a, b, c$ (Methods), as given by

$$I(Ca_{res}; \Delta t) \simeq a \log_2(b + c \cdot V). \tag{22}$$

This function fitted well the volume-dependent mutual information (Fig. 6a, red line) and the mutual information per PF input (Fig. 6b, red line), indicating that this function captures the features of the volume-dependency of the mutual information in the spine volume and the larger volume.

We also considered the Gaussian channel, simple linear transmission system. For the input $X$, when the system noise $Z$ obeys the Gaussian distribution, the output $Y = X + Z$ also obeys the Gaussian distribution. In this case, under the constraint $E[X^2] < F$, the mutual information (channel capacity) between the input, $X$, and the output, $Y$, is simply described as

$$I(Y; X) = \frac{1}{2} \log_2\left(1 + \frac{F}{\sigma_Z^2}\right), \tag{23}$$

where, $F$ denotes the power constraint of input, and $\sigma_Z$ denotes the standard deviation of the noise intensity. Here, $F$ is regarded as a constant value because the input distribution for calculating $I(Ca_{res}; \Delta t)$ in the equation (22) (Fig. 6, blue line) is assumed to be unchanged. It has been shown that the standard deviation of reactions is proportional to the power of number of the molecules, *i.e.* volume, so that the fluctuation of number of molecules can be approximated as $\sigma_Z'^2 \propto V$ [24].

Then, the fluctuation of concentration of molecules can be approximated by $\sigma_Z^2 = (\sigma'_Z/V)^2 \propto V^{-1}$. Therefore, the mutual information for the Gaussian channel is given by

$$I(Ca_{res}; \Delta t) \simeq \frac{1}{2} \log_2(1 + c \cdot V) \qquad (24)$$

(Fig. 6a, b, blue lines). Equations (22) and (24) indicates the same volume-dependency of the mutual information. However, in the smaller volume including the spine volume, the mutual information per PF input of the $Ca^{2+}$ response was larger than that of the Gaussian channel and the fitted function (Fig. 6b). This difference in volume-dependency is likely to be caused by the different values of the parameters $a = 0.3924651$ and $c = 0.5128671$, which were larger than $1/2$ and $1$ in the fitted function and equal in the Gaussian channel, respectively. There were other differences in both systems; the noise of the system in this study is not exactly a Gaussian noise, and the input-output relation is nonlinear. Despite such differences, both systems exhibited the similar volume-dependency of the mutual information, suggesting that the more efficient information transfer in the smaller volume is a universal property in the general information transduction systems.

**Discussion**

In this study, we made the simple stochastic model of $Ca^{2+}$ increase in the spine of PF-cerebellar Purkinje cell synapse. We clarified the mechanisms of the robustness, sensitivity, and efficiency, and showed that these properties become prominent in the spine volume (Fig. 7). The robustness appears under the condition where the standard deviation of the distribution of the $Ca^{2+}$ response, intrinsic noise, is larger than the fluctuation of the distribution of the $Ca^{2+}$ response caused by the PF input fluctuation, extrinsic noise.

The higher sensitivity to the lower amplitude of PF input requires the wider dynamic range of $Ca^{2+}$ response and the smaller standard deviation of the distribution of $Ca^{2+}$ response in the range of the lower amplitude of PF input. In the spine volume, because of the stochastic facilitation caused by the stochasticity in reactions [25], even the weak PF input can induce a large $Ca^{2+}$ increase, resulting in a wider dynamic range of $Ca^{2+}$ response in the range of the lower amplitude of PF input in the spine volume than in the cell volume. Moreover, the standard deviation of the distribution of the $Ca^{2+}$ response in the range of the lower amplitude of PF input was small. In the larger volume than the spine volume, the sensitivity abruptly decreases because stronger PF input was required for compensating the large standard deviation of the distribution of $Ca^{2+}$ response.

The highest efficiency in the spine volume is derived from the nature of the volume-dependency of mutual information; the rate of increase of the mutual information monotonically decreased with the increase in volume. Then, the mutual information per PF input, efficiency, becomes larger in the smaller volume. This result indicates that the spine utilizes the limit of the smallness to acquire the highest efficiency.

Robustness appears when intrinsic noise is larger than extrinsic noise. Sensitivity appears because of the stochastic facilitation. Efficiency appears because of the nature of volume-dependency of information transfer. These characteristics are derived from the smallness, which we denote "the small-volume effect". The small-volume effect enables the spine robust, sensitive and efficient information transfer. The small-volume effect may be seen not only in spines, but also in other small intracellular organelles, and general strategy for biological information transfer. The small-volume effect is one of the reasons why the spine has to be so small. The small-volume effect is also equivalent to the small-number effect, suggesting that the robustness, sensitivity, and efficiency can also be seen under the conditions where numbers of molecules are limited even in a larger volume.

It has been known that in most of the excitable neurons, the $Ca^{2+}$ increase in the spine by the glutamine stimulus depends mainly on NMDAR, another glutamate-gated $Ca^{2+}$ channel [26–28]. In the future, we will analyze whether the $Ca^{2+}$ increase by NMDAR in the spine also shows the robustness, sensitivity and efficiency and ask whether such properties are conserved among the spines regardless of the types of the receptors.

**Methods**

**Simple stochastic model**

The block diagram of the simple stochastic model is the same as that of the simple deterministic model (Fig. 1e) [1]. The inputs are $PF$ and $CF$. After the Fig. 3 unless specified in Fig. 6, we set $CF = 0$, and used only $PF$ as the input. The output is $Ca$.

The total cytosolic $Ca^{2+}$ in the spine of the Purkinje cell, $Ca$, is derived from the three pathways.

$$Ca = Ca_{basal} + Ca_{VGCC} + Ca_{IP_3}. \qquad (25)$$

where $Ca_{basal}$, $Ca_{VGCC}$ and $Ca_{IP_3}$ denote the basal cytosolic $Ca^{2+}$, $Ca^{2+}$ through voltage-gated $Ca^{2+}$ channel (VGCC) triggered by $CF$, and through IP$_3$ receptors of the internal $Ca^{2+}$ store triggered by $PF$, respectively. $Ca_{basal}$ is constantly produced and described by

$$\phi \xrightarrow{C_b/\tau_{FB}} Ca_{basal} \xrightarrow{1/\tau_{FB}} \phi, \qquad (26)$$

where $C_b/\tau_{FB}$ and $1/\tau_{FB}$ denote the production and decay rate constants of $Ca_{basal}$, respectively. Hereafter, $\phi$ denotes a fixed value.

$Ca_{VGCC}$ is triggered by $CF$, and described by

$$CF \xrightarrow{1/\tau_{CF}} Ca_{VGCC} \xrightarrow{1/\tau_{CF}} \phi, \tag{27}$$

where $1/\tau_{CF}$ denotes the production and decay rate constants of $Ca_{VGCC}$, respectively. $CF$ is given by $Amp_{CF} \cdot V$ at $t = tCF$.

$Ca_{IP_3}$ is produced as follows. Briefly, $PF$ produces $IP_3$. $Ca$ has a positive feedback ($FB$) through the activation of IP$_3$ receptor ($G_{IP_3R}$). $IP_3$ and $G_{IP_3R}$ synergistically induce $Ca$ release through IP$_3$R ($Ca_{IP_3}$). $IP_3$ is triggered by $PF$, and described by

$$PF \xrightarrow{1/\tau_{PF}} IP_3 \xrightarrow{1/\tau_{PF}} \phi, \tag{28}$$

where $1/\tau_{PF}$ denotes the production and decay rate constants of $IP_3$. $PF$ is given by $Amp_{PF} \cdot V$ at $t = t_{PF}$.

The time-delay variable $FB$ is described by

$$Ca \xrightarrow{1/\tau_{FB}} FB \xrightarrow{1/\tau_{FB}} \phi, \tag{29}$$

where $1/\tau_{FB}$ denotes the production and decay rate constants of $FB$. This decay rate constant also determines the degradation rate of $Ca_{IP_3}$.

IP$_3$ receptor coupled with Ca$^{2+}$, $G_{IP_3R}$, is mediated by the positive and negative feedbacks from $FB$, and is given by the nonlinear function, described by

$$G_{IP_3R} = Amp_{G_{IP_3R}} \left\{ \frac{k \cdot FB}{(k + FB)(K + FB)} \right\}^{n_{G_{IP_3R}}}, \tag{30}$$

where $Amp_{G_{IP_3R}}$, $k$, $K$ and $n_{G_{IP_3R}}$ denote the amplitude of feedback, thresholds of $FB$ for the positive and negative feedbacks, and non-linearity of feedback, respectively.

$Ca$ released from $IP_3R$, $Ca_{IP_3}$, is described by

$$Ca_{IP_3} = IP_3 \cdot \frac{G_{IP_3R}}{V}. \tag{31}$$

These reactions are simulated by the use of Gillespie's method and τ-leap method [29]. The τ-leap method by Cao *et al.* shows good approximation for the first-order reactions like this study [30].

We defined the $Ca_{res}$ as the AUC of the time course of Ca$^{2+}$, given by

$$Ca_{res} = \int_T \{[Ca](t) - C_b\} dt, \tag{32}$$

where $C_b$ denotes the basal concentration of Ca$^{2+}$, which is 41.6 nM. Note that Koumura et al.[22] defined $Ca_{res}$ as the logarithmic AUC, which is different from that in this study, however, the results of this study qualitatively show the same results.

The values of the parameters in the simple stochastic model are shown in Supplementary Table 1 and 2. The parameters excluding the follows are the same as the simple deterministic model [22].

**Mutual information between the PF- and CF-timing and Ca$^{2+}$ increase**

We measured the input timing information coded by the Ca$^{2+}$ response by mutual information between the $Ca_{res}$ and PF-CF timing interval $\Delta t = t_{CF} - t_{PF}$, given by

$$I(Ca_{res}; \Delta t) = \int_{\Delta t} p_{in}(\tau) \left( \int_{Ca_{res}} p_c(c|\tau) \log_2 \frac{p_c(c|\tau)}{\int_{\Delta t} p_{in}(\tau) p_c(c|\tau) \, d\tau} dc \right) d\tau. \tag{33}$$

Here, the $p_{in}(\Delta t)$ follows the uniform distribution. To remove the bias by the bin width of $Ca_{res}$, the mutual information was calculated by the method introduced by Cheong et al. [31]. The mutual information remains the almost constant for the bin width between 10$^{-2}$ and 10$^{-3.5}$ μm$^3$. Therefore, we fixed the bin width of $Ca_{res}$ as 10$^{-2}$ [pM·min] in the analysis and drawing the histogram.

We also measured the information coded by the probability of the large Ca$^{2+}$ increase and by the amplitude of the Ca$^{2+}$ increase, denoted as the mutual information of the probability component and of the amplitude component, respectively. We defined $\theta$ as the $Ca_{res}$ representing the local minimum value of the marginal distribution $p_c(Ca_{res})$ (Fig. 1) and $s$ as the logical value whether $Ca_{res} > \theta$ is satisfied or not.

The mutual information coded of the probability component is defined as

$$I_{prob}(Ca_{res}; \Delta t) = \int_{\Delta t} p_{in}(\tau) \left( \int_{Ca_{res}} p_c(c|\tau) \log_2 \frac{p_c(c|\tau)}{p_{-prob}(c|\tau)} dc \right) d\tau,$$

$$p_{-prob}(Ca_{res}|\Delta t) = \sum_{s \in \{0,1\}} P(s) p_c(Ca_{res}|s, \Delta t), \tag{34}$$

where $p_{-prob}(Ca_{res}|\Delta t)$ denotes the distribution of $Ca_{res}$ without the probability component, which was calculated by marginalizing $\Delta t$ out of the probability component $p_c(s|\Delta t)$ in $p_c(Ca_{res}|\Delta t)$.

The mutual information coded by the amplitude component is defined as

$$I_{amp}(Ca_{res}; \Delta t) = \int_{\Delta t} p_{in}(\tau) \left( \int_{Ca_{res}} p_c(c|\tau) \log_2 \frac{p_c(c|\tau)}{p_{-amp}(c|\tau)} dc \right) d\tau,$$

$$p_{-amp}(Ca_{res}|\Delta t) = \sum_{s \in \{0,1\}} P(s|\Delta t) p_c(Ca_{res}|s), \tag{35}$$

where $p_{-amp}(Ca_{res}|\Delta t)$ denotes the distribution of $Ca_{res}$ without the amplitude component, which was calculated by marginalizing $\Delta t$ out of the amplitude component $p_c(Ca_{res}|s, \Delta t)$ in $p_c(Ca_{res}|\Delta t)$. These informations satisfies

$$I(Ca_{res}; \Delta t) = I_{prob}(Ca_{res}; \Delta t) + I_{amp}(Ca_{res}; \Delta t). \tag{36}$$

**Mutual information between the amplitude of PF input and Ca$^{2+}$ increase**

We also calculated the mutual information between $Ca_{res}$ and $Amp_{PF}$ assuming the input distribution as the Gaussian distribution with $\mu_s$, the average, and $STD$, the standard deviation, given by

$$I(Ca_{res}; Amp_{PF}) = \int_{Amp_{PF}} p_s(a|\mu_s, STD) \left( \int_{Ca_{res}} p_c(c|a) \log_2 \frac{p_c(c|a)}{\int_{Amp_{PF}} p_s(a|\mu_s, STD) p_c(c|a) \, da} dc \right) da, \tag{37}$$

where

$$p_s(Amp_{PF}|\mu_s, STD) = \frac{1}{\sqrt{2\pi STD^2}} \exp\left[-\frac{(Amp_{PF} - \mu_s)^2}{2STD^2}\right]. \tag{38}$$

Note that $p_s$ and $p_a$ have $Amp_{PF}$ as the variable and are assumed as a Gaussian distribution. However, these distributions mean different feature. $p_s$ means the input distribution of the mutual information, and $p_a$ means the distribution of amplitude of PF input under the fluctuation of amplitude of PF input.

**Fitted function of the volume-dependency of mutual information**

The mutual information per PF input against the volume was fitted by the functions, $a \log_2(b + cV)/V$ and $1/2 \log_2(1 + cV)/V$ (Fig. 6b), by use of the Nonlinear Least Squares method with the Marquadt-Levenberg algorithm. We obtained fitting line, $a \log_2(b + cV)$, with the best fit $a = 0.3924651$, $b = 1.049141$, $c = 1.330285$, and the channel capacity of the Gaussian channel, $1/2 \log_2(1 + cV)$, with the best fit parameter $c = 0.5128671$.

**Acknowledgement**

We are grateful to Dr. Hidetoshi Urakubo, Dr. Takuya Koumura, Dr. Shinsuke Uda, Dr. Yuichi Sakumura and our laboratory members for their critical reading of this manuscript, and Dr. Tamiki Komatsuzaki for fruitful discussion. This work was supported by The Creation of Fundamental Technologies for Understanding and Control of Biosystem Dynamics, CREST, from the Japan Science and Technology (JST).

**Additional information**

**Author contributions**: M.F. and S.K. conceived the project. M.F. constructed the model and performed the stochastic simulation. M.F., K.O. Y.K. M.H. and S.K. analyzed the data. M.F and K.O. contributed theoretical analysis. M.F. and S.K. wrote the manuscript.

**Competing financial interests**: The authors declare that they have no conflict of interest.

**References**

1. Napper, R. M. & Harvey, R. J. Number of parallel fiber synapses on an individual Purkinje cell in the cerebellum of the rat. *J. Comp. Neurol.* **274,** 168–77 (1988).

2. Stuart, G., Spruston, N. & Häusser, M. *Dendrites*. (Oxford University Press, 2007). doi:10.1093/acprof:oso/9780198566564.001.0001

3. Harris, K. M. & Stevens, J. K. Dendritic spines of rat cerebellar Purkinje cells: serial electron microscopy with reference to their biophysical characteristics. *J. Neurosci.* **8,** 4455–69 (1988).

4. Rapp, M., Segev, I. & Yarom, Y. Physiology, morphology and detailed passive models of guinea-pig cerebellar Purkinje cells. *J. Physiol.* **474,** 101–18 (1994).

5. Takács, J. & Hámori, J. Developmental dynamics of Purkinje cells and dendritic spines in rat cerebellar cortex. *J. Neurosci. Res.* **38,** 515–30 (1994).

6. Antunes, G. & De Schutter, E. A stochastic signaling network mediates the probabilistic induction of cerebellar long-term depression. *J. Neurosci.* **32,** 9288–300 (2012).

7. Anwar, H., Hepburn, I., Nedelescu, H., Chen, W. & De Schutter, E. Stochastic calcium mechanisms cause dendritic calcium spike variability. *J. Neurosci.* **33,** 15848–67 (2013).

8. Barbour, B. Synaptic currents evoked in Purkinje cells by stimulating individual granule cells. *Neuron* **11,** 759–69 (1993).

9.  Okubo, Y. *et al.* Imaging extrasynaptic glutamate dynamics in the brain. *Proc. Natl. Acad. Sci. U. S. A.* **107,** 6526–31 (2010).

10. Isope, P. & Barbour, B. Properties of unitary granule cell→Purkinje cell synapses in adult rat cerebellar slices. *J. Neurosci.* **22,** 9668–78 (2002).

11. Ito, M. Neurophysiological aspects of the cerebellar motor control system. *Int. J. Neurol.* **7,** 162–76 (1970).

12. Kawato, M. Internal models for motor control and trajectory planning. *Curr. Opin. Neurobiol.* **9,** 718–27 (1999).

13. Kawato, M., Kuroda, S. & Schweighofer, N. Cerebellar supervised learning revisited: biophysical modeling and degrees-of-freedom control. *Curr. Opin. Neurobiol.* **21,** 791–800 (2011).

14. Miyakawa, H., Lev-Ram, V., Lasser-Ross, N. & Ross, W. N. Calcium transients evoked by climbing fiber and parallel fiber synaptic inputs in guinea pig cerebellar Purkinje neurons. *J. Neurophysiol.* **68,** 1178–89 (1992).

15. Wang, S. S., Denk, W. & Häusser, M. Coincidence detection in single dendritic spines mediated by calcium release. *Nat. Neurosci.* **3,** 1266–73 (2000).

16. Ito, M. The molecular organization of cerebellar long-term depression. *Nat. Rev. Neurosci.* **3,** 896–902 (2002).

17. Mauk, M. D., Garcia, K. S., Medina, J. F. & Steele, P. M. Does cerebellar LTD mediate motor learning? Toward a resolution without a smoking gun. *Neuron* **20,** 359–62 (1998).

18. Steinmetz, J. E. Classical nictitating membrane conditioning in rabbits with varying interstimulus intervals and direct activation of cerebellar mossy fibers as the CS. *Behav. Brain Res.* **38,** 97–108 (1990).

19. Doi, T., Kuroda, S., Michikawa, T. & Kawato, M. Inositol 1,4,5-trisphosphate-dependent Ca2+ threshold dynamics detect spike timing in cerebellar Purkinje cells. *J. Neurosci.* **25,** 950–61 (2005).

20. Honda, M., Urakubo, H., Koumura, T. & Kuroda, S. A common framework of signal processing in the induction of cerebellar LTD and cortical STDP. *Neural Netw.* **43,** 114–24 (2013).

21. Honda, M., Urakubo, H., Tanaka, K. & Kuroda, S. Analysis of development of direction selectivity in retinotectum by a neural circuit model with spike timing-dependent plasticity. *J. Neurosci.* **31,** 1516–27 (2011).

22. Koumura, T., Urakubo, H., Ohashi, K., Fujii, M. & Kuroda, S. Stochasticity in Ca2+ increase in spines enables robust and sensitive information coding. *PLoS One* **9,** e99040 (2014).

23. Rancz, E. A. & Häusser, M. Dendritic calcium spikes are tunable triggers of cannabinoid release and short-term synaptic plasticity in cerebellar Purkinje neurons. *J. Neurosci.* **26,** 5428–37 (2006).

24. Jin Wang & Peter Wolynes. Intermittency of single molecule reaction dynamics in fluctuating environments. *Phys. Rev. Lett.* **74,** 4317–4320 (1995).

25. McDonnell, M. D. & Ward, L. M. The benefits of noise in neural systems: bridging theory and experiment. *Nat. Rev. Neurosci.* **12,** 415–426 (2011).

26. Garaschuk, O., Schneggenburger, R., Schirra, C., Tempia, F. & Konnerth, A. Fractional Ca2+ currents through somatic and dendritic glutamate receptor channels of rat hippocampal CA1 pyramidal neurones. *J. Physiol.* **491,** 757–772 (1996).

27. Kubota, S. & Kitajima, T. A model for synaptic development regulated by NMDA receptor subunit expression. *J. Comput. Neurosci.* **24,** 1–20 (2008).

28. Lester, R. A. J. & Jahr, C. E. NMDA channel behavior depends on agonist affinity. *J. Neurosci.* **12,** 635–643 (1992).

29. Cao, Y., Gillespie, D. T. & Petzold, L. R. Efficient step size selection for the tau-leaping simulation method. *J. Chem. Phys.* **124,** 044109 (2006).

30. Gillespie, D. T., Hellander, A. & Petzold, L. R. Perspective: Stochastic algorithms for chemical kinetics. *J. Chem. Phys.* **138,** 170901 (2013).

31. Cheong, R., Rhee, A., Wang, C. J., Nemenman, I. & Levchenko, A. Information Transduction Capacity of Noisy Biochemical Signaling Networks. *Science (80-. ).* **334,** 354–358 (2011).

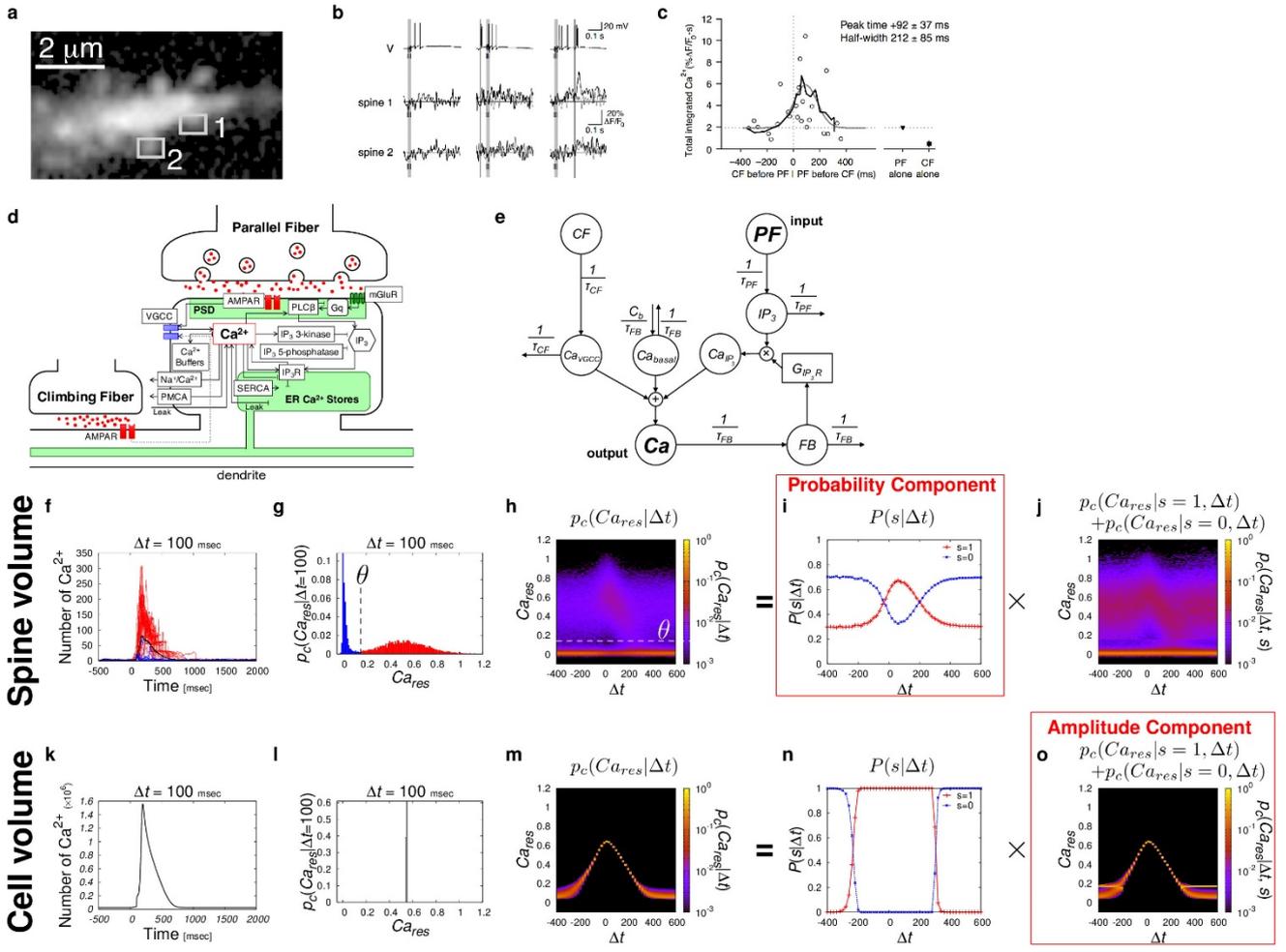

**Figure 1 | Information transfer of PF- and CF-timing by probability of Ca$^{2+}$ increase in the simple stochastic model.** (a-c) Experimental results of Ca$^{2+}$ increase by PF and CF inputs at the spine in the cerebellar Purkinje cell [15]. (a) Spines of the cerebellar Purkinje cell. (b) Ca$^{2+}$ response in the indicated spines in (a). V indicates membrane potential and ΔF/F0 indicates the normalized changes of the fluorescence probe of Ca$^{2+}$. The left, middle, and right panels show the time courses with only the PF input (shaded vertical line), with the CF input (black vertical line) before 60 msec before the PF input, and with the PF input 60 msec before the CF input, respectively. (c) Total integrated Ca$^{2+}$ with PF and CF inputs with various timing. The gray line indicates the best fits of the raw data points to Gaussian functions. The black line indicates the box-smoothed average over three points. (d) Schematic representation of Ca$^{2+}$ increase by PF and CF inputs in the detailed stochastic model [22]. (e) The block diagram of the simple stochastic model in this study (see Methods). After Fig. 3, we set $CF = 0$, and used only $PF$ as the input. Ca$^{2+}$ increase in the spine volume ($10^{-1}$ μm$^3$) (f–j) in the cell volume ($10^3$ μm$^3$) (k–o) in the simple stochastic model. (f, k), Ca$^{2+}$ increase with $\Delta t = 100$ msec. $\Delta t$ indicates the timing interval between PF and CF inputs, which is set the timing of PF input as 0, and $\Delta t$ with the PF input before CF input is positive and *vice versa*. In (f), the large Ca$^{2+}$ increase (red) and small Ca$^{2+}$ increase (blue) divided by $\theta$ in (g). (g, l), the probability

density distribution of $Ca_{res}$. $Ca_{res}$ denotes the area under the curve of the time course of $Ca^{2+}$ shown in (f and k). In (g), the threshold $\theta$ is defined as the local minimum of the marginal distribution for $\Delta t$, given by $p_c(Ca_{res}) = \int_{\Delta t} p_c(Ca_{res}|\tau) p_{in}(\tau) d\tau$ (see Supplementary Fig. 1). (h, m), the probability density distribution of $Ca_{res}$ in the spine volume (h) and cell volume (m). (i, n), the probability component of the distribution of $Ca_{res}$ that exceeds the threshold $\theta$ in the spine volume ($s = 1$) (see Methods). Since the distribution of $Ca_{res}$ in the cell volume is unimodal distribution, for convenience, we set $\theta = 0.157$ in the cell volume, the same as that in the spine volume. (j, o), the amplitude component of the distribution of $Ca_{res}$ (see Methods)

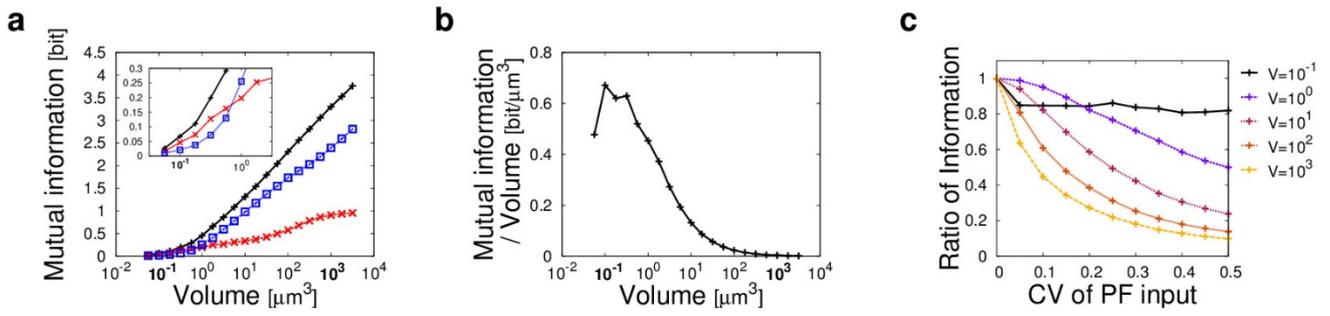

**Figure 2 | The efficient and robust features in the simple stochastic model.** (a) The volume-dependency of the mutual information between the PF- and CF-timing and $Ca_{res}$. The black, red, blue lines indicate the mutual information of the total distribution of $Ca_{res}$, of the probability component and of the amplitude component, respectively. (b) The volume-dependency of the mutual information per volume. (c) CV of amplitude of PF input-dependency of the mutual information. The ratio of information was obtained by setting mutual information with CV = 0 for each volume at 1.

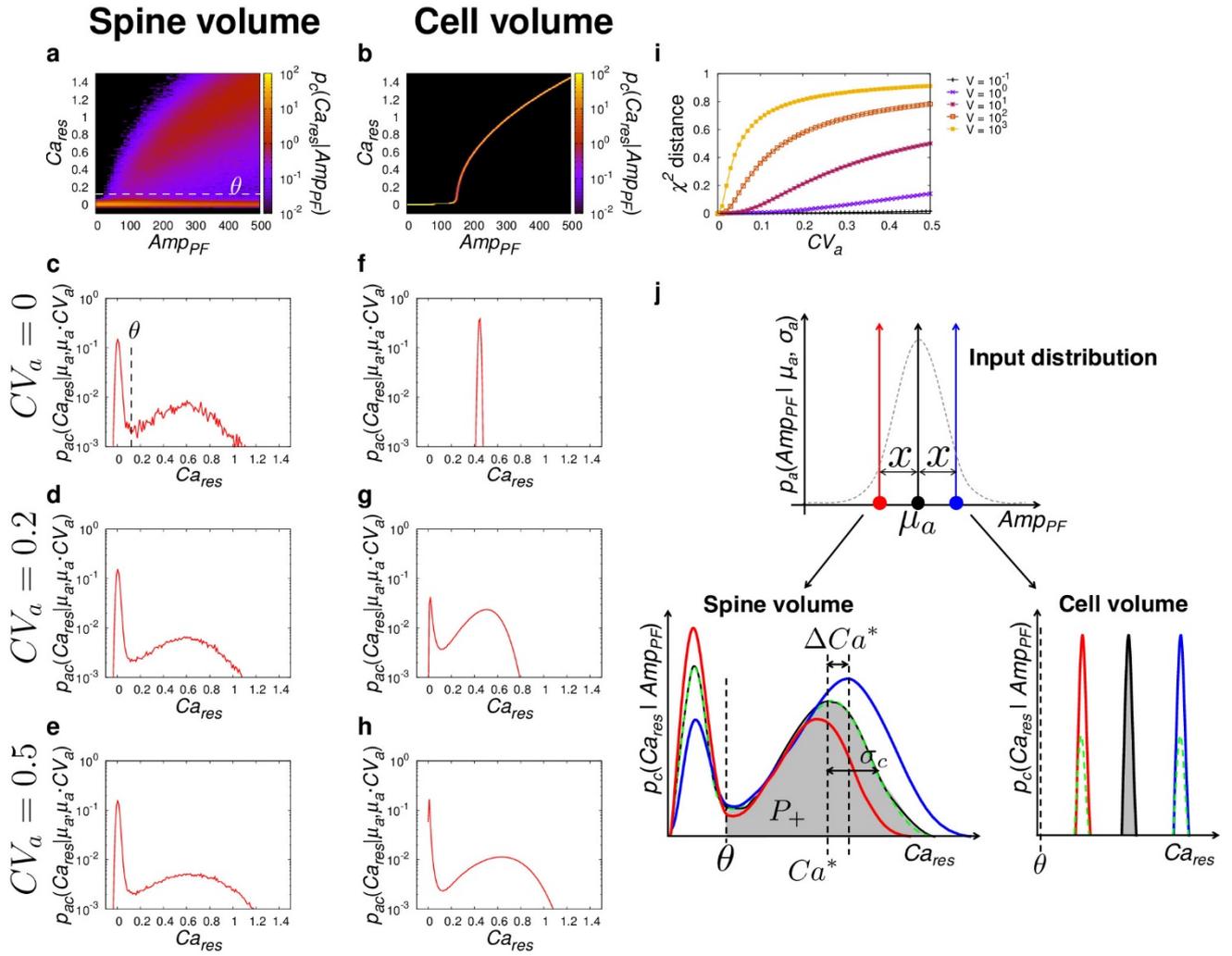

**Figure 3 | Unchanging of distributions of $Ca_{res}$ against fluctuation of PF input gives robustness in the spine volume.** The $Amp_{PF}$-dependency of $p_c(Ca_{res}|Amp_{PF})$, the distributions of $Ca_{res}$, in the spine volume (a) and in the cell volume (b). (c–h) The distributions of $Ca_{res}$ against the PF input fluctuation with the indicated $CV_a$ with $\mu_a = 180$ in the spine volume (c–e) and in the cell volume (f–h), respectively. $\theta$ indicates the threshold dividing the distribution into the ranges with large $Ca_{res}$ and with small $Ca_{res}$ (see Supplementary Fig. 1). (i) The $CV_a$-dependency of $\chi^2$ distance of distributions between $p_{ac}(Ca_{res}|\mu_a, \mu_a \cdot CV_a)$ and $p_{ac}(Ca_{res}|\mu_a, 0)$ with $\mu_a = 180$. (j) The input distribution of $Amp_{PF}$ is given by the Gaussian distribution $\mathcal{N}(\mu_a, \sigma_a^2)$. The distribution of $Ca_{res}$ with $Amp_{PF} = \mu_a$ (black), $Amp_{PF} = \mu_a + x$ (blue), and $Amp_{PF} = \mu_a - x$ (red) in the spine volume (left) and the cell volume (right). The averaged distribution of the distributions with $Amp_{PF} = \mu_a + x$ and with $Amp_{PF} = \mu_a - x$ (green).

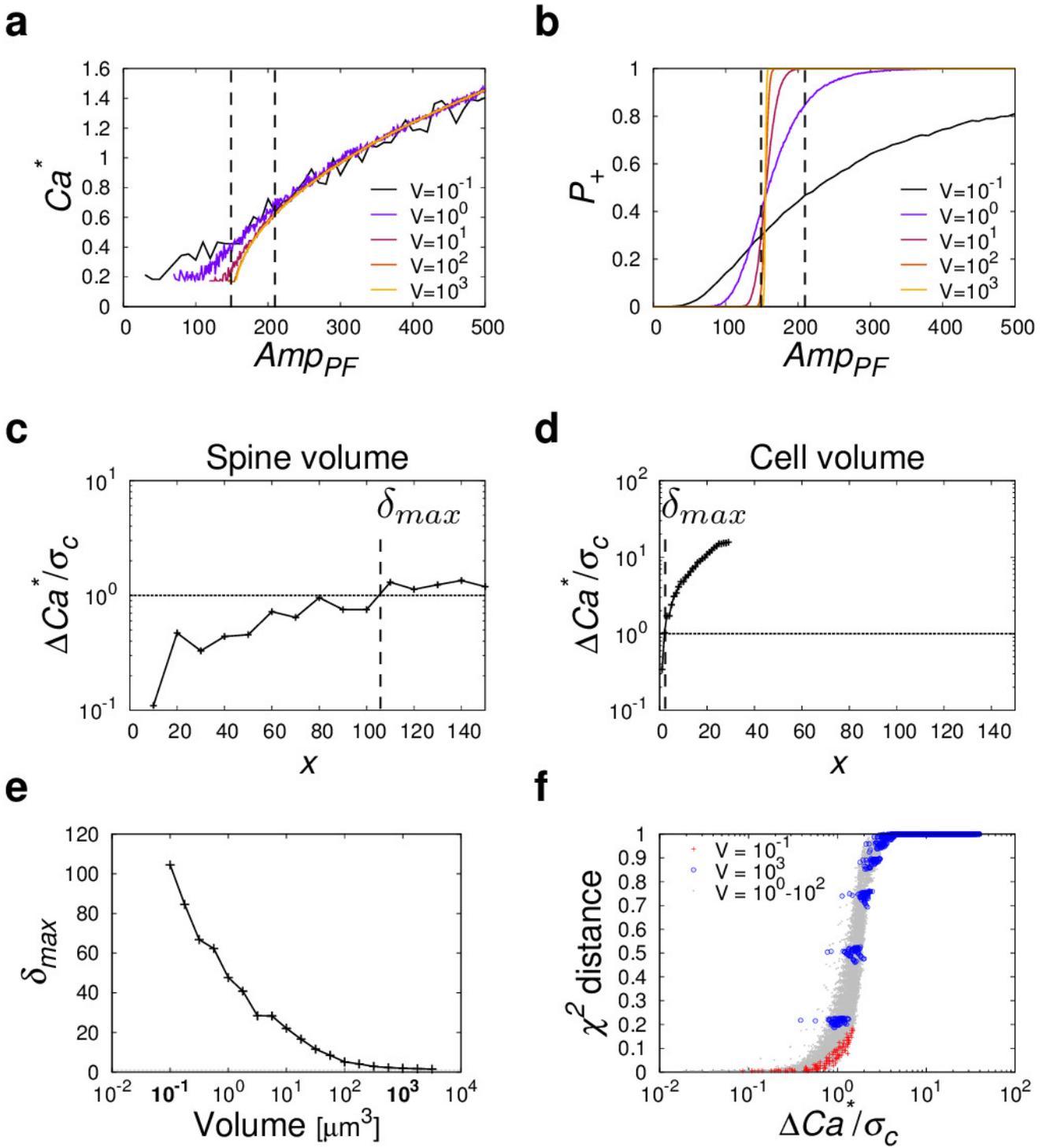

**Figure 4 | The ratio between extrinsic and intrinsic noise, $\Delta Ca^*/\sigma_c$, determines the range of the robustness.** (a) The $Amp_{PF}$-dependency of the $Ca_{res}$ realizing the peak of distribution of $Ca_{res}$ for $Ca_{res} > \theta$, $Ca^*$. (b) The $Amp_{PF}$-dependency of the probability of $Ca_{res} > \theta$, $P_+$. (c, d) $Amp'_{PF} = x$, the relative amplitude of PF input,-dependency of $\Delta Ca^*/\sigma_c$ with $\mu_a = 180$ in the spine volume (c) and the cell volume (d). The robustness index $\delta_{max}$ is defined as $x$ giving $\Delta Ca^*/\sigma_c = 1$ (e) The volume-dependency of $\delta_{max}$ for $\mu_a = 180$. (f) The relationship between $\Delta Ca^*/\sigma_c$ and the $\chi^2$ distance

between the averaged distribution of the distributions of $Ca_{res}$ with $Amp'_{PF} = \pm x$ and the distribution of $Ca_{res}$ with $Amp'_{PF} = 0$. The red points, blue points, and gray dots indicate the value obtained in the spine volume, the cell volume and the intermediate volumes, respectively.

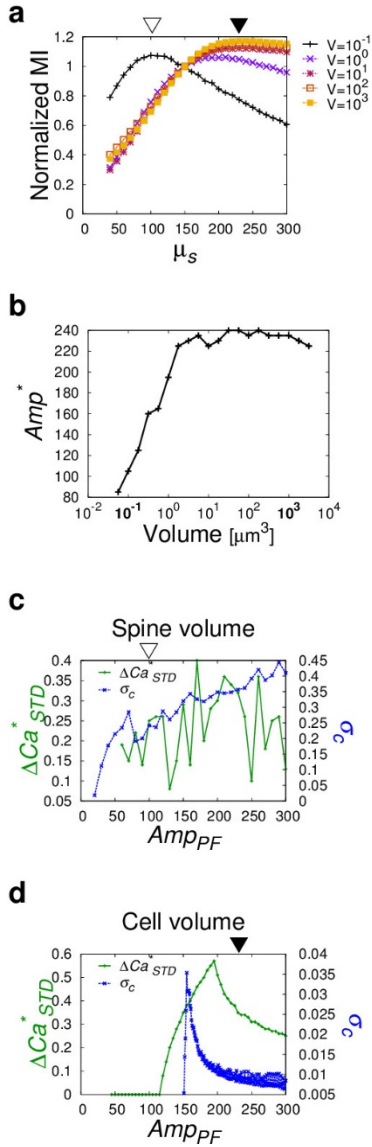

**Figure 5 | The mechanism of the sensitivity.** (a) $\mu_s$, the average of input distribution of $Amp_{PF}$,-dependency of the mutual information. The mutual information is normalized by the value of that with $\mu_s = 150$. The brighter color indicates the larger volume. The white and black triangles denote $Amp^*$, the amplitude of the PF input realizing the maximum of the mutual information, in the spine volume and in the cell volume, respectively. $Amp^* = 100$ in the spine volume, and $Amp^* = 235$ in the cell volume. The input distribution of $Amp_{PF}$ is utilized as the Gaussian distribution with the standard deviation, $STD$, 40. (b) The volume-dependency of the amplitude of the PF input realizing the maximum of the mutual information, $Amp^*$. (c, d) The $Amp_{PF}$-dependencies of $\Delta Ca^*_{STD}$, the dynamic range of the distribution of $Ca_{res}$ for $Ca_{res} > \theta$ (green), and $\sigma_c$, the standard deviation of the distribution of $Ca_{res}$ (blue). $STD = 40$ was used. The white and black triangles denote $Amp^*$ in the spine volume and in the cell volume, respectively.

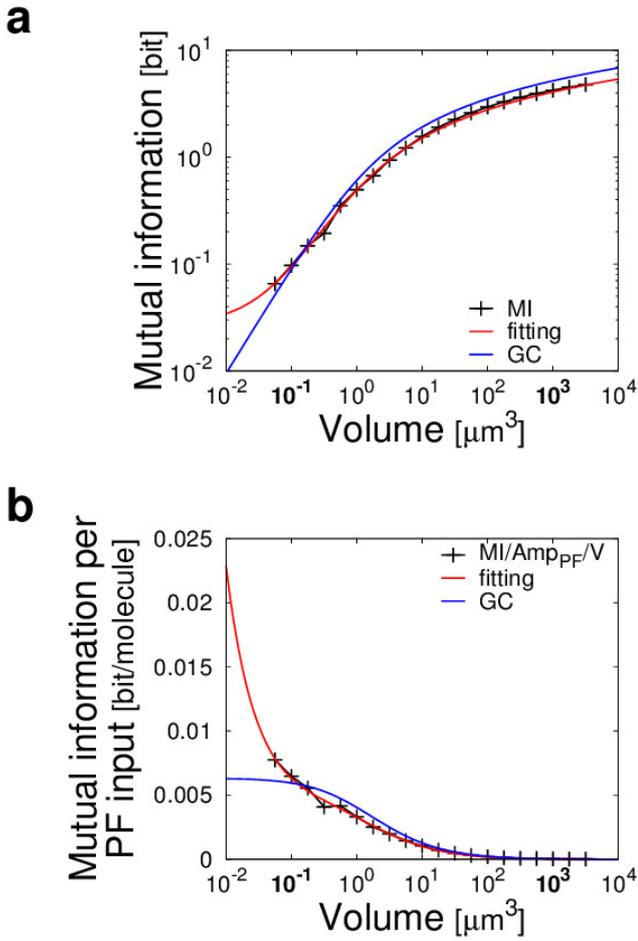

**Figure 6 | The mechanism of the efficiency.** (a) The volume-dependency of the mutual information between $Amp_{PF}$ and $Ca_{res}$. (b) The volume-dependency of the mutual information per PF input, *i.e.* the efficiency. The total mutual information, black; the fitted curve of the total mutual information by $a \log_2(b + cV)$ with $a = 0.3924651$, $b = 1.049141$, $c = 1.330285$, red; the channel capacity of the Gaussian channel, $1/2 \log_2(1 + cV)$ with $c = 0.5128671$, blue (see Methods). We assume the input distribution of $Amp_{PF}$ as the Gaussian distribution with $\mu_s = 150$, the average of amplitude of the distribution of the PF input, and $STD = 40$, the standard deviation of the distribution of the PF input.

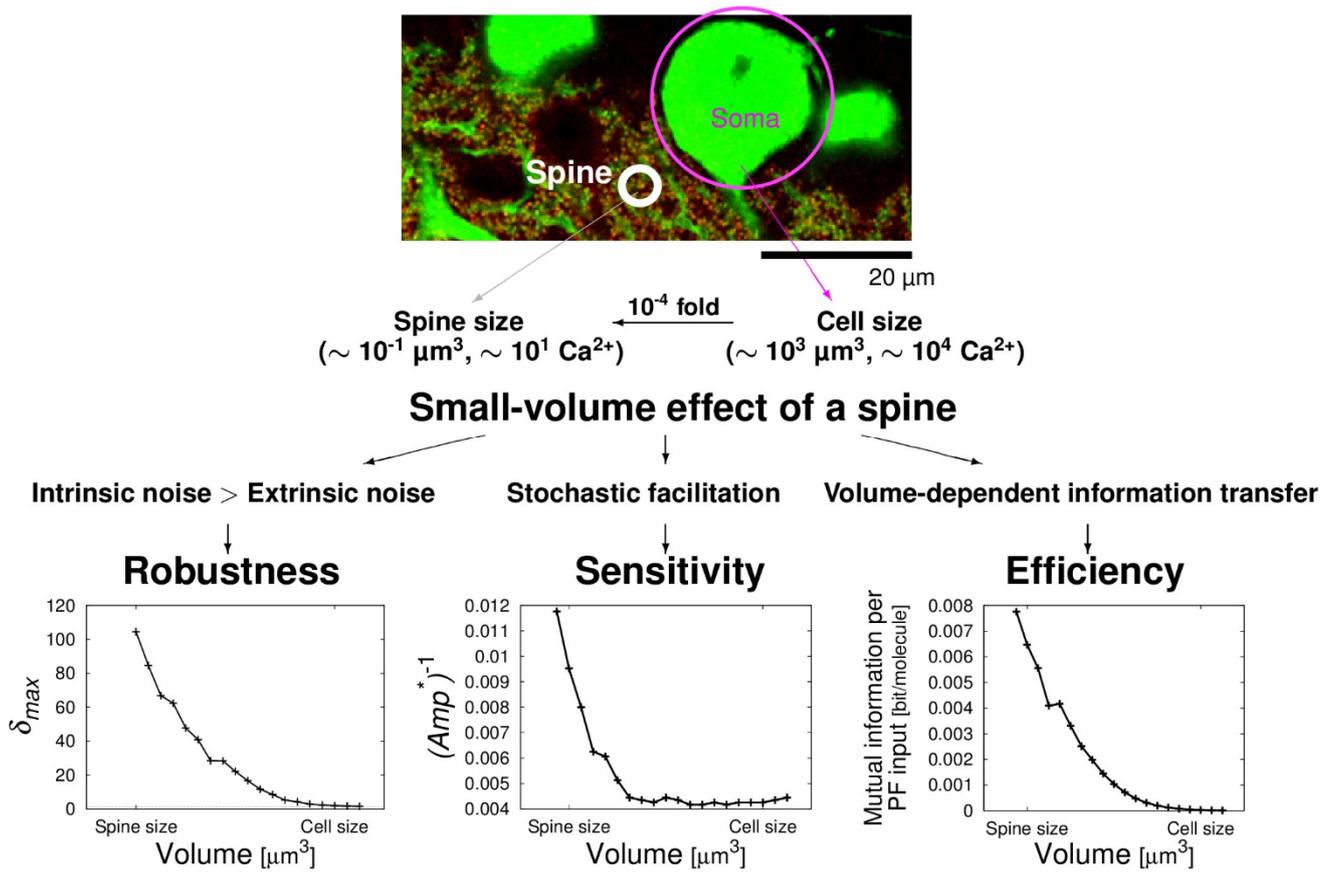

**Figure 7 | Summarizing figure.** The small-volume effect enables the spine robust, sensitive and efficient information transfer. Robustness appears when intrinsic noise is larger than extrinsic noise. Sensitivity appears because of the stochastic facilitation. Note that, as index for the sensitivity in this figure, the inverse of $Amp^*$ (Fig. 5b) is used. Efficiency appears because of the nature of volume-dependency of information transfer.

**Supplementary Information**

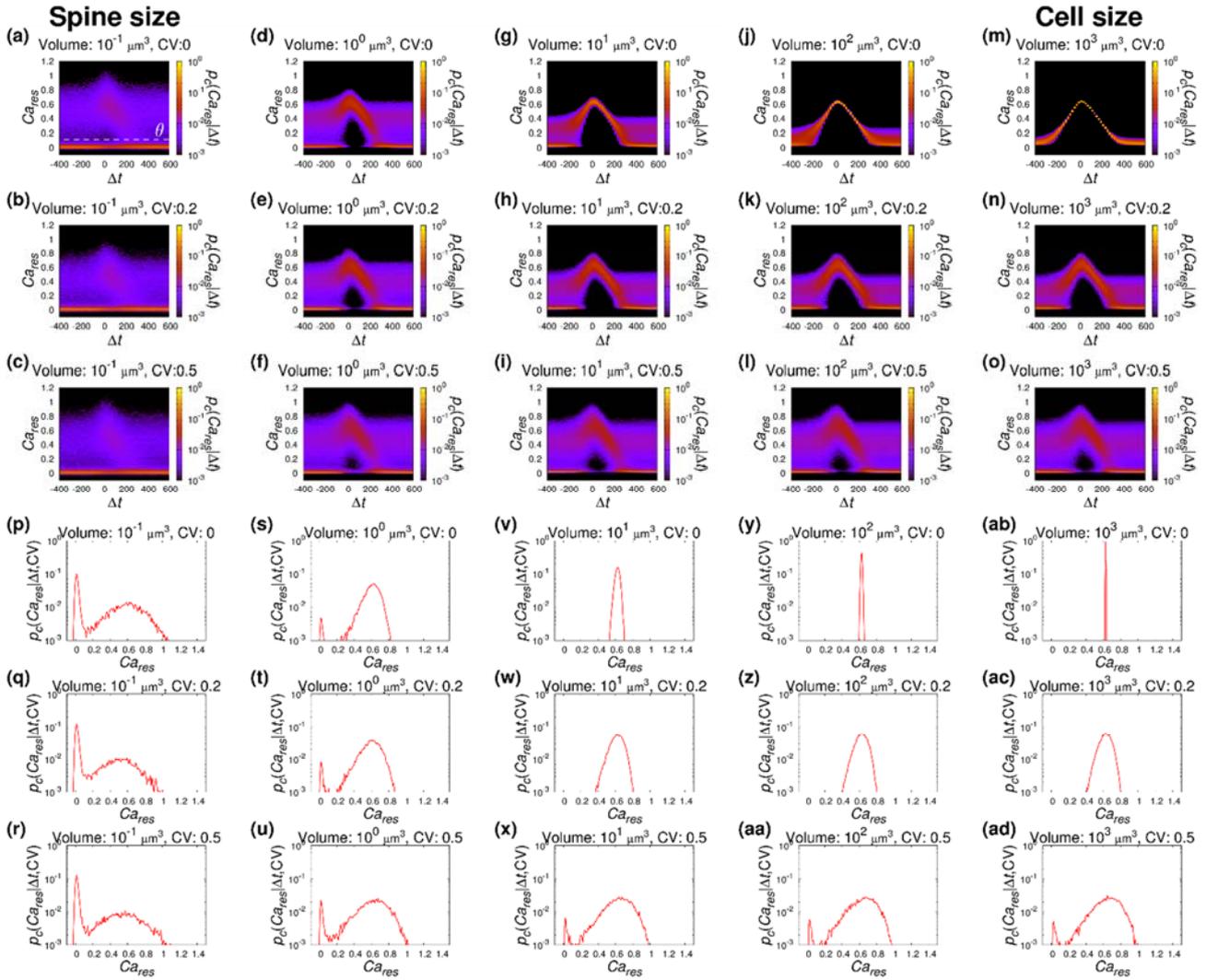

**Supplementary Figure 1** | (a–o) The $\Delta t$-dependency of the distribution of $Ca_{res}$. The volume and the CV of amplitude of PF input are indicated. In the spine volume, the distribution of $Ca_{res}$ is divided into two distributions by threshold $\theta = 0.157$ defined as the local minimum of the marginal distribution of $Ca_{res}$ for $\Delta t$ s.t. $p_c(Ca_{res}) = \int_{\Delta t} p_{in}(\tau) p_c(Ca_{res}|\tau) \, d\tau$. (p–ad) The cross sections of (a–o) with $\Delta t = 0$. This distribution of $Ca_{res}$ in the spine volume remained the same regardless of the $CV_a$ value, whereas, that in the cell volume largely varied.

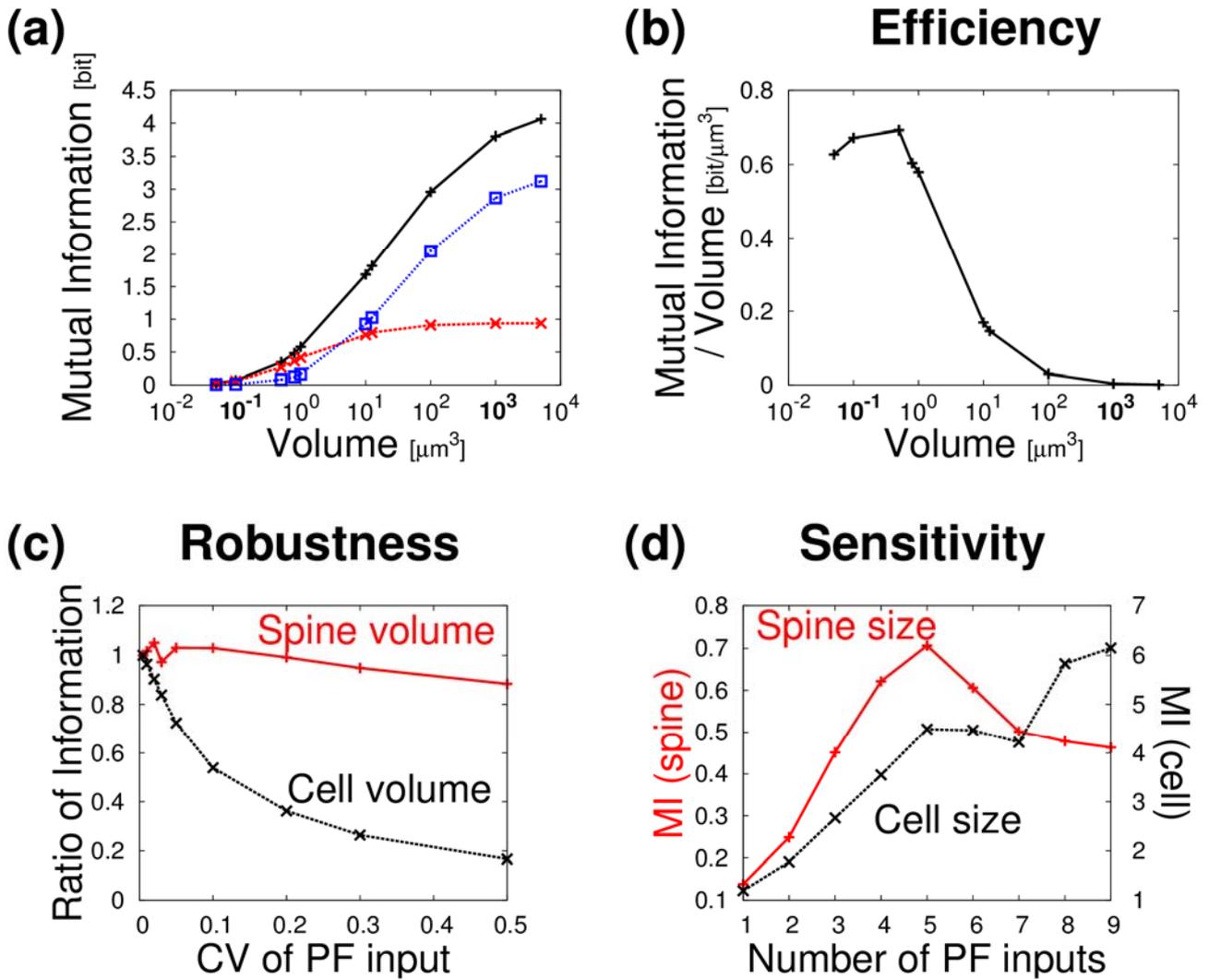

**Supplementary Figure 2** | The efficient, robust and sensitive features of $Ca^{2+}$ increase in the previous study using the detailed stochastic model [22]. (a) The volume-dependency of the mutual information between the PF- and CF-timing, $\Delta t$, and $Ca^{2+}$ response, $Ca_{res}$. The total mutual information, black; that of the probability component, red; that of the amplitude component, blue. (b) The volume-dependency of the mutual information per volume. (c) The CV of amplitude of PF input-dependency of the mutual information. (d) The number of PF inputs-dependency of the mutual information. In the detailed stochastic model, we denote that the spine volume is $10^{-1}$ μm$^3$ and the cell volume is $5 \times 10^3$ μm$^3$ [22].

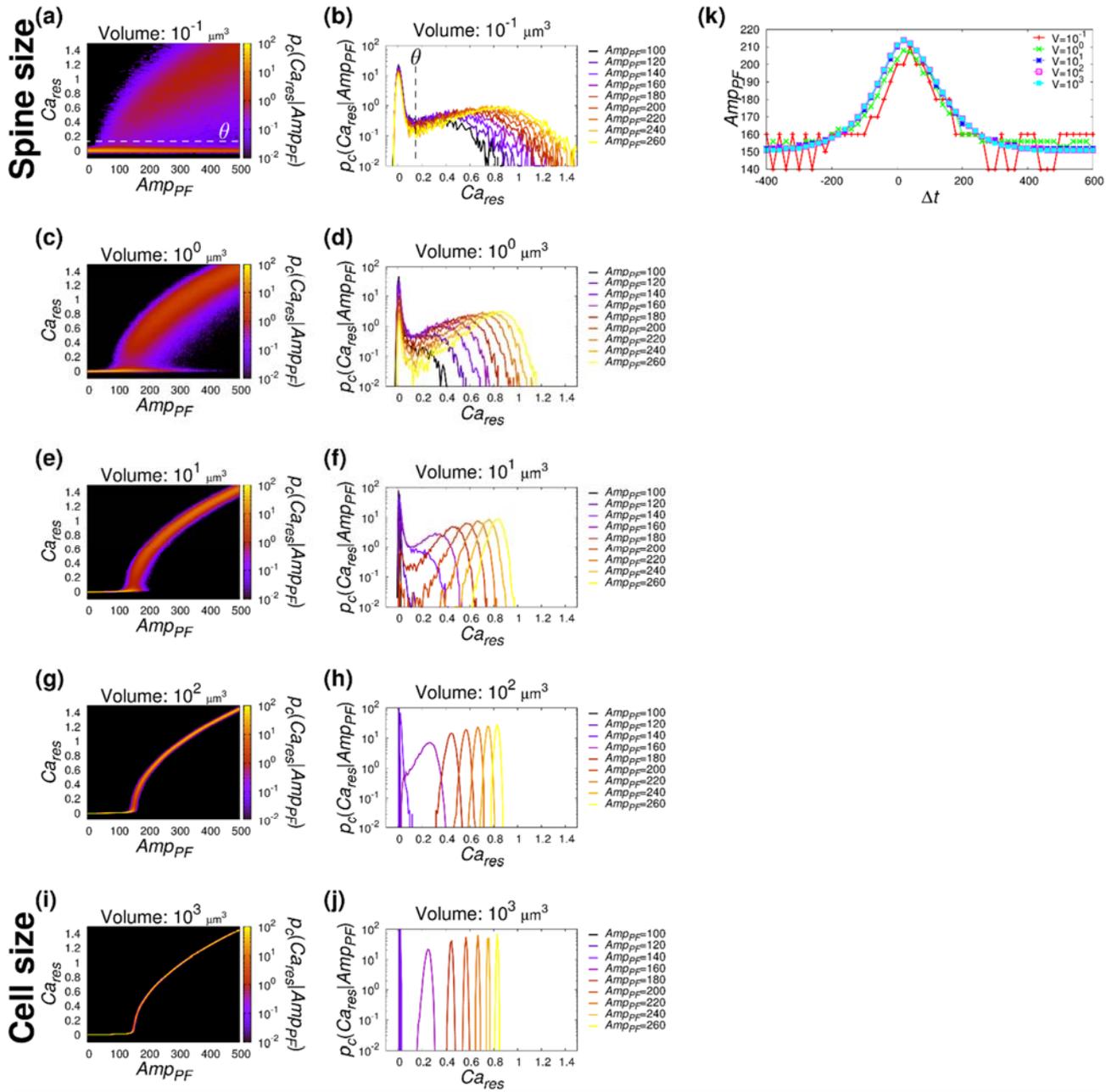

**Supplementary Figure 3** | The $Amp_{PF}$-dependency of the distribution of $Ca_{res}$ in the indicated volumes. (a, c, e, g, i) The distribution of $Ca_{res}$. (b, d, f, h, j) The cross section of distribution of $Ca_{res}$ at the indicated $Amp_{PF}$. $\theta (=0.157)$ indicates the threshold dividing the distribution into the ranges with large $Ca_{res}$ and with small $Ca_{res}$ (see Supplementary Fig. 1). (k) The $Amp_{PF}$ that gives $p_c(Ca_{res}|Amp_{PF})$, the distribution of $Ca_{res}$ with PF input alone, closest to $p_c(Ca_{res}|\Delta t)$, the distribution of $Ca_{res}$ with PF and CF inputs with various $\Delta t$.

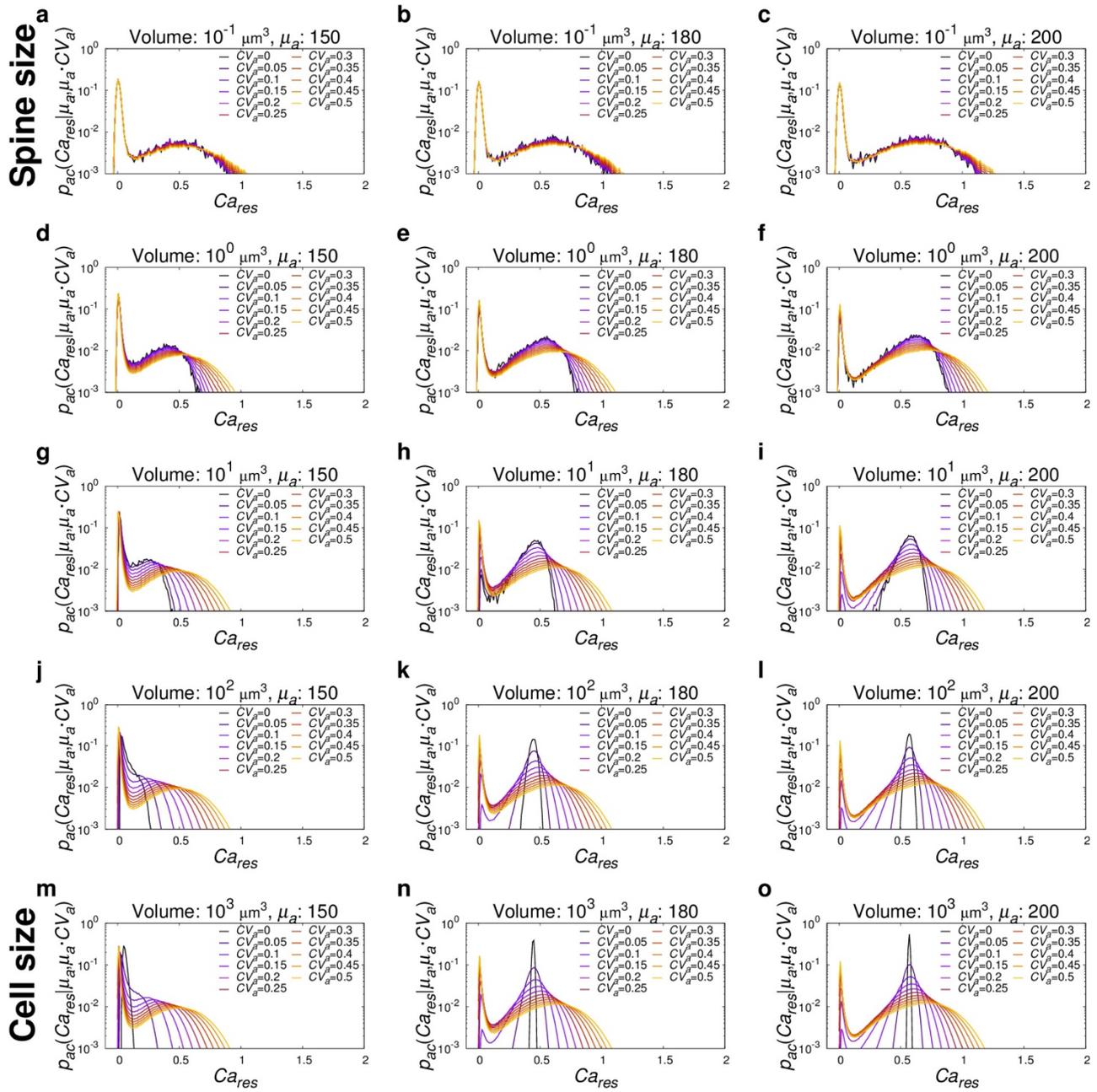

**Supplementary Figure 4** | The distributions of $Ca_{res}$ against $CV_a$ with the indicated volumes and the average of $Amp_{PF}$, $\mu_a$.

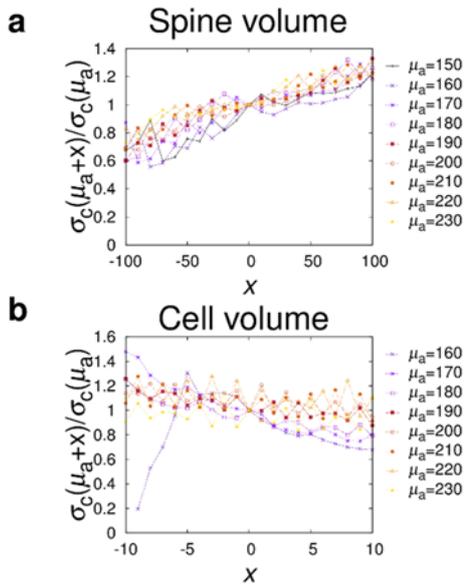

**Supplementary Figure 5** | $\sigma_c(\mu_a + x)$ can be regarded as $\sigma_c(\mu_a)$ up to the upper bound of the range of $x$ satisfying the equation (4). (a) the spine volume. (b) the cell volume. $\sigma_c(\mu_a + x)/\sigma_c(\mu_a)$ were almost within the range between 0.8 and 1.2, assuming that $\sigma_c(\mu_a + x)$ is approximated by $\sigma_c(\mu_a)$. The upper bound of the range of $x$ satisfying the equation (4) in the spine and cell volumes are determined by $\delta_{\max}$ (see Fig. 4c, d).

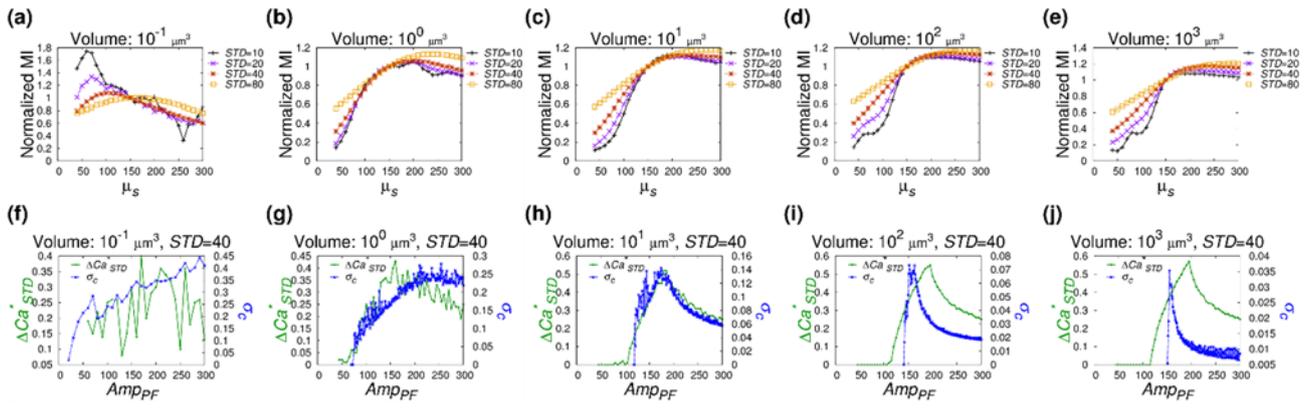

**Supplementary Figure 6** | The mechanism of the sensitivity. (a–e) $\mu_s$, the average of input distribution of $Amp_{PF}$,-dependency of the mutual information normalized by the value of that with $Amp_{PF} = 150$. (f–j) The $Amp_{PF}$-dependencies of the dynamic range of the distribution of $Ca_{res}$ for $Ca_{res} > \theta$. The volume is indicated.

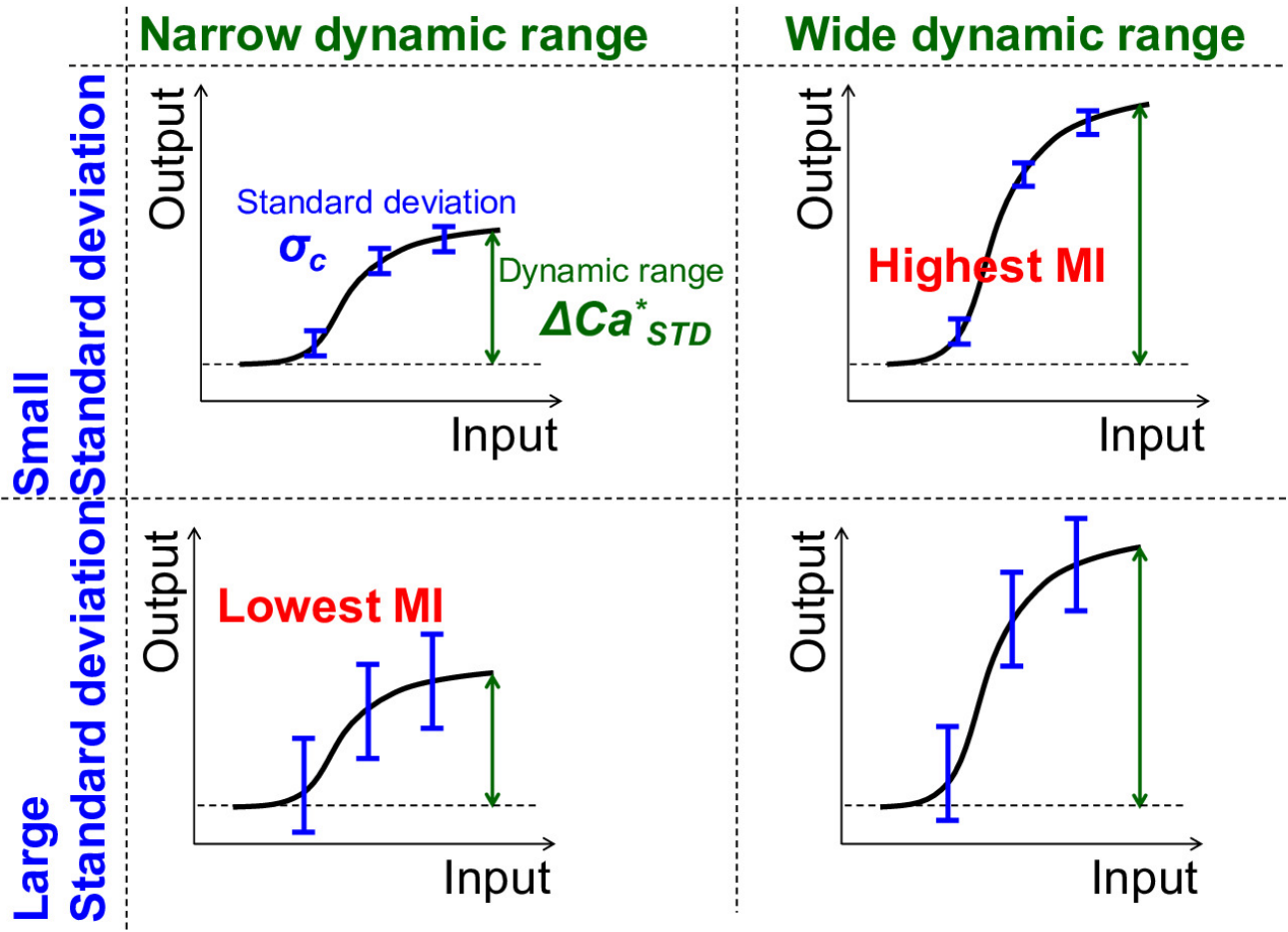

**Supplementary Figure 7** | The mutual information depends on both $\Delta Ca^*_{STD}$, the dynamic range, and $\sigma_c$, the standard deviation of the distribution of the output. In general, if the input distribution is the same, the wider $\Delta Ca^*_{STD}$, the dynamic range of the output, gives the higher mutual information when $\sigma_c$, the standard deviation of the output, is the same (compare the left and right panels). The smaller $\sigma_c$ gives the higher mutual information when the $\Delta Ca^*_{STD}$ is the same (compare the top and bottom panels).

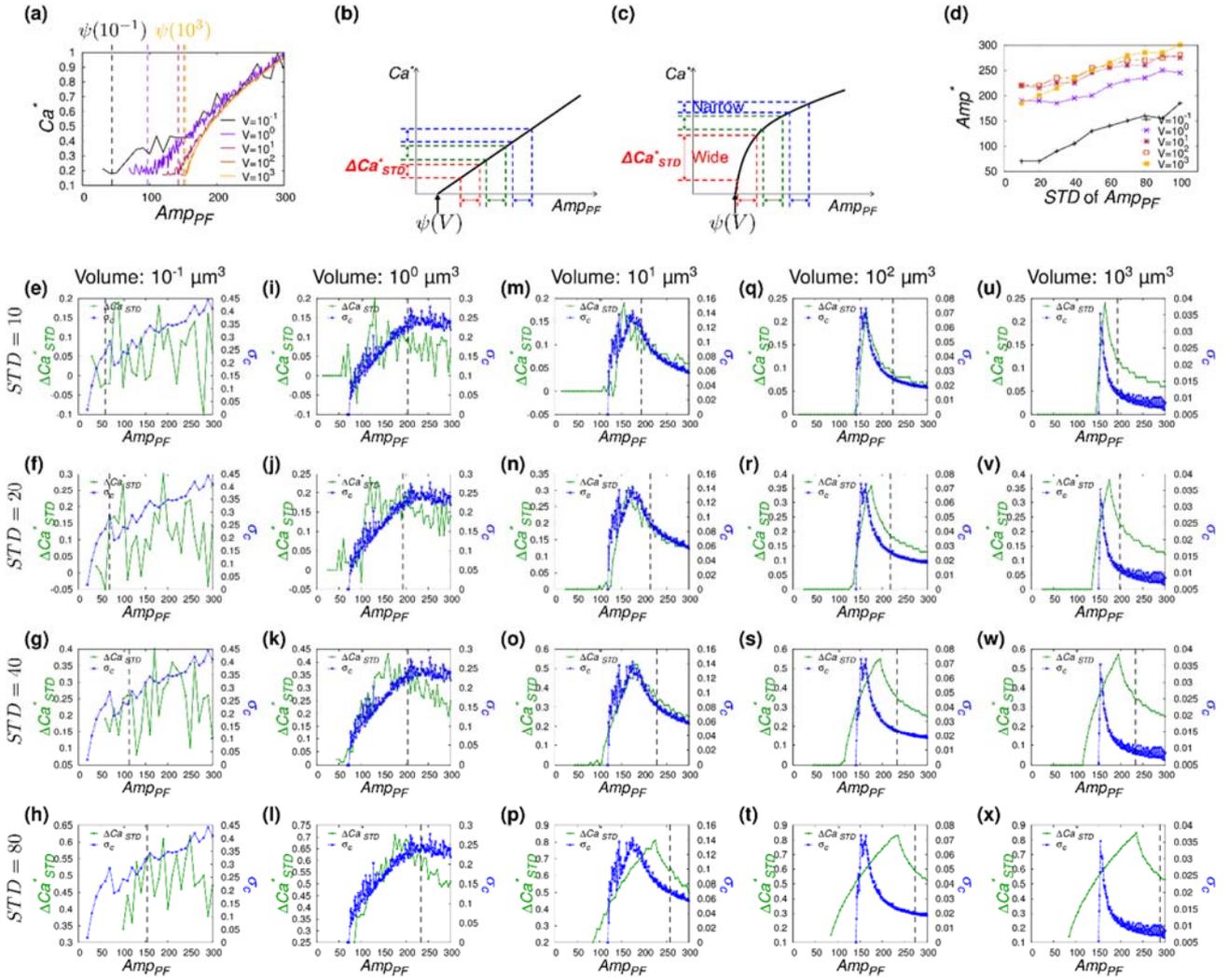

**Supplementary Figure 8** | $\Delta Ca^*_{STD}$, the dynamic range, and $\sigma_c$, the standard deviation of the distribution of the output. (a) The $Amp_{PF}$-dependency of $Ca^*$, $Ca_{res}$ realizing the peak of the distribution of $Ca_{res}$ for $Ca_{res} > \theta$. We defined $\psi(V)$ for each volume as the $Amp_{PF}$ where the $Ca^*$ begins to increase. In the spine volume, $\psi(10^{-1})$ was around 50, whereas, $\psi(10^3)$ was around 150 in the cell volume. (b, c) The schematic representation of the relationship between $Amp_{PF}$ and $Ca^*$ in the spine volume (b) and in the cell volume (c). (d) The $STD$ of $Amp_{PF}$-dependency of $Amp_{PF}$ giving the maximum of the mutual information, $Amp^*$. (e–x) The $Amp_{PF}$-dependencies of $\Delta Ca^*_{STD}$, the dynamic range of the distribution of $Ca_{res}$ for $Ca_{res} > \theta$, (green) and $\sigma_c$, the standard deviation of the distribution of $Ca_{res}$ (blue). The volume and $STD$ are indicated.

**Supplementary Table 1 | Parameters of the simple stochastic model in this study**

| Parameters | Values |
|---|---|
| $\tau_{PF}$ [msec] | 120 |
| $\tau_{CF}$ [msec] | 10 |
| $\tau_{FB}$ [msec] | 80 |
| $Amp_{G_{IP_3R}}$ | 1291.6667 |
| $k$ [1/μm³] | 626.3027 |
| $K$ [1/μm³] | 626.3027 |
| $n_{IP_3R}$ | 2.7 |
| $C_b$ [1/μm³] | 27.70185 |
| $V$ [μm³] | $10^{-1}$–$10^{3}$ |

**Supplementary Table 2 | Parameters which are different between the cases with various PF- and CF-timing and with single PF input alone.** Note that the simple deterministic model showed the same results as that of the detailed deterministic model; however, by the reduction of the model, the PF and CF inputs were non-dimensional values. By the loss of the dimension of the number of molecules, we could not perform the stochastic simulation as it is. Therefore, we re-determined the numbers of PF and CF inputs as follows. For the PF input, the number of PF input becomes smaller than 1 in the spine volume (0.1 μm³), but the number of PF input needs to be the positive integer. We increased the PF input to 6-fold from the simple deterministic model so that the amount of IP$_3$, the mediator of PF input, is the same as that in the detailed stochastic model, resulting in the amplitude of a PF input in the spine volume as 3 ($Amp_{PF} \times V = 30.11 \times 0.1 = 3.011 \simeq 3$). We reduced the reaction rate constant of the Ca$^{2+}$ release by the binding IP$_3$ and IP$_3$R to one sixth to compensate the $Ca_{IP_3}$. For the CF input, the CF input increased to 6-fold so that the number of Ca$^{2+}$ through the CF input in the simple stochastic model becomes the same amount as that in the detailed stochastic model.

| Parameters | Values | |
|---|---|---|
| | PF and CF input (Figs. 1, 2) | PF input alone (Figs. 3, 4, 5, 6) |
| $Amp_{CF}$ [1/μm³] | 361.328 | None |
| $Amp_{PF}$ [1/μm³] | 30.11×5 times | variable×1 time |
| $t_{CF}$ [msec] | variable | None |
| $t_{PF}$ [msec] | {0, 10, 20, 30, 40} | 0 |
| CV of PF input | variable | 0 (in simulation) |

**Supplementary Note 1 | Proof. The distribution of $Ca_{res}$ changes with the fluctuation of $Amp_{PF}$ when the equation (5) in the main text is satisfied.**

There does not exist the distribution of $Ca_{res}$ with the fluctuation of $Amp_{PF}$, $p_{ac}(Ca_{res}|\mu_a, \sigma_a)$, which does not change against the fluctuation of $Amp_{PF}$ under the conditions where the equation (5) in the main text is satisfied.

When the distribution of $Ca_{res}$ with $Amp_{PF} = \mu_a$, $p_c(Ca_{res}|\mu_a)$, is not the same as the averaged distribution of $Ca_{res}$ between with $Amp_{PF} = \mu_a + (a - \mu_a) = a$ and with $Amp_{PF} = \mu_a - (a - \mu_a) = 2\mu_a - a$, $1/2[p_c(Ca_{res}|a) + p_c(Ca_{res}|2\mu_a - a)]$, i.e.

$$p_c(Ca_{res}|\mu_a) \neq \frac{1}{2}[p_c(Ca_{res}|a) + p_c(Ca_{res}|2\mu_a - a)]. \tag{1}$$

By the reductio ad absurdum, we demonstrate that $CV_a^* = \sigma_a^*/\mu_a$ of $Amp_{PF}$ satisfying the following equation for $0 < \sigma_a < \sigma_a^*$ corresponding to $0 < CV_a < CV_a^*$ does not exist.

$$p_c(Ca_{res}|\mu_a) \simeq p_{ac}(Ca|\mu_a, \sigma_a) = \int_{\mu_a}^{\infty} \mathcal{N}(a|\mu_a, \sigma_a^2)[p_c(Ca_{res}|a) + p_c(Ca_{res}|2\mu_a - a)]\, da$$

$$\Leftrightarrow \int_{\mu_a}^{\infty} \{\mathcal{N}(a|\mu_a, \sigma_a^2)[p_c(Ca_{res}|a) + p_c(Ca_{res}|2\mu_a - a)] - 2p_c(Ca_{res}|\mu_a)\}\, da \simeq 0$$

$$\int_{\mu_a}^{\infty} \mathcal{N}(a|\mu_a, \sigma_a^2)[p_c(Ca_{res}|a) + p_c(Ca_{res}|2\mu_a - a) - 2p_c(Ca_{res}|\mu_a)]\, da \simeq 0. \tag{2}$$

This equation means that the distribution of $Ca_{res}$ does not change against the fluctuation of $Amp_{PF}$. For simplicity, $f(a)$ is defined as

$$f(a) \equiv p_c(Ca_{res}|a) + p_c(Ca_{res}|2\mu_a - a) - 2p_c(Ca_{res}|\mu_a). \tag{3}$$

The equation (2) becomes

$$\int_{\mu_a}^{\infty} \mathcal{N}(a|\mu_a, \sigma_a^2) f(a)\, da \simeq 0$$

$$a \geq \mu_a, f(\mu_a) = 0, f(a) \not\equiv 0, f(a): \text{Continuous}. \tag{4}$$

Note that from the equation (1), $f(a) \not\equiv 0$. Here, $g(\sigma_a)$ is defined as

$$g(\sigma_a) = \int_{\mu_a}^{\infty} \mathcal{N}(a|\mu_a, \sigma_a^2) f(a)\, da, \tag{5}$$

and it is assumed that there exists $\sigma_a$ satisfying $g(\sigma_a) \equiv 0$ for $0 \leq \sigma_a \leq \sigma_a^*$. From $f(a) \not\equiv 0$ and $f(a)$ is the continuous function, there exists $a^*$ satisfying $f(a) \lessgtr 0$ and $f(a^*) = 0$ for $0 < a < a^*$ (Supplementary Fig. 9, upper panel). Then, we obtained

$$\left|\int_{\mu_a}^{a^*} \mathcal{N}(a|\mu_a, \sigma_a^2)f(a)\, da\right| = \left|\int_{a^*}^{\infty} \mathcal{N}(a|\mu_a, \sigma_a^2)f(a)\, da\right|. \tag{6}$$

This means that the integral of $\mathcal{N}(a|\mu_a, \sigma_a^2)f(a)$ in the range of $0 < a < a^*$ is the same as that in the range of $a > a^*$ from $g(\sigma_a^*) = 0$. In the followings, for the simple explanation, we assume $f(a) > 0$ for $0 < a < a^*$. Note that the same explanation can be provided when $f(a) < 0$ for $0 < a < a^*$. If $g(\sigma_a)$ is always 0 regardless $\sigma_a$, differential of $g(\sigma_a)$ must be 0. Then, we differentiates $g(\sigma_a)$ by $\sigma_a$, given by

$$\frac{dg(\sigma_a)}{d\sigma_a} = \frac{d}{d\sigma_a}\int_{\mu_a}^{\infty} \mathcal{N}(a|\mu_a, \sigma_a^2)f(a)\, da = \int_{\mu_a}^{\infty} \frac{d}{d\sigma_a}\mathcal{N}(a|\mu_a, \sigma_a^2)f(a)\, da$$
$$= \int_{\mu_a}^{\infty} \frac{(a-\mu_a)^2 - \sigma_a^2}{\sigma_a^3}\mathcal{N}(a|\mu_a, \sigma_a^2)f(a)\, da \tag{7}$$

$\mathcal{N}(a|\mu_a, \sigma_a^2)f(a)$ for $a < \mu_a + \sigma_a$ becomes far from 0 with the increase in $\sigma_a$, and that for $a > \mu_a + \sigma_a$ becomes close to 0 (Supplementary Fig. 9, middle panel). Then, when $\sigma_a$ slightly decreases from $\sigma_a = a^* - \mu_a$, $\int_{\mu_a}^{a^*} \mathcal{N}(a|\mu_a, \sigma_a^2)f(a)\, da$ becomes larger and $\int_{a^*}^{\infty} \mathcal{N}(a|\mu_a, \sigma_a^2)f(a)\, da$ becomes smaller. Hence, the l.h.s. of the equation (6) becomes larger and the r.h.s. of the equation (6) becomes smaller, then, $g(\sigma_a)$ increases. When $\sigma_a$ decreases more, $\int_{\mu_a}^{a^*} \mathcal{N}(a|\mu_a, \sigma_a^2)f(a)\, da$ keeps positive value, and $\int_{a^*}^{\infty} \mathcal{N}(a|\mu_a, \sigma_a^2)f(a)\, da$ decreases and becomes close to 0, then, the summation of $\int_{\mu_a}^{a^*} \mathcal{N}(a|\mu_a, \sigma_a^2)f(a)\, da$ and $\int_{a^*}^{\infty} \mathcal{N}(a|\mu_a, \sigma_a^2)f(a)\, da$, i.e. $g(a)$ becomes positive value (Supplementary Fig. 9, lower panel). This means there does not exist $\sigma_a$ satisfying $g(\sigma_a) \equiv 0$ for $0 < \sigma_a < a^* - \mu_a$. This is to be conflict with the assumption that there exists $\sigma_a$ satisfying $g(\sigma_a) = 0$ for $0 \leq \sigma_a \leq \sigma_a^*$. Then, there does not exist $\sigma_a$ satisfying $g(\sigma_a) = 0$ for $0 \leq \sigma_a \leq \sigma_a^*$. Therefore, if the equation (1) is satisfied, the equation (2) becomes not to be satisfied when $CV_a$ slightly becomes large. Thus, under the conditions where the equation (5) in the main text is satisfied, there does not exist the distribution of $Ca_{res}$ with the fluctuation of $Amp_{PF}$, $p_{ac}(Ca_{res}|\mu_a, \sigma_a)$, which does not change against the fluctuation of $Amp_{PF}$, indicating that the distribution of $Ca_{res}$ changes with the increase in $CV_a$.

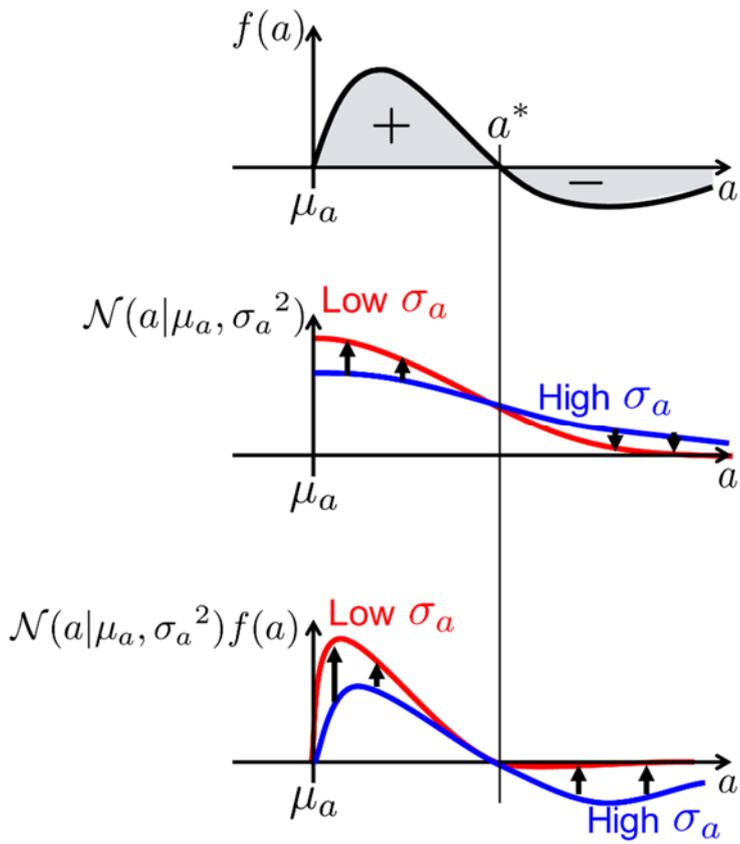

**Supplementary Figure S9** | The schematic image of $\sigma_a$-dependency of $f(a)$ (upper panel), $\mathcal{N}(a|\mu_a, \sigma_a)$ (middle panel) and $\mathcal{N}(a|\mu_a, \sigma_a)f(a)$ (lower panel). When $\sigma_a$ decreases from $\sigma_a = a^*$, $\mathcal{N}(a|\mu_a, \sigma_a)$ increases for $0 < a < a^*$ and decreases for $a > a^*$. Therefore, the integral of $N(a|\mu_a, \sigma_a)f(a)$ for $0 < a < a^*$ increases and that for $a > a^*$ becomes close to 0.